\documentclass[12pt,table]{article}
\usepackage{amsfonts}

\usepackage{textalpha} %

\usepackage{bm}
\usepackage{enumerate} 
\usepackage{amssymb,amsmath}
\usepackage{amsthm} 
\usepackage{caption}
\usepackage{subcaption}
\usepackage{accents} 

\usepackage{dsfont}

\usepackage{stackengine}

\usepackage{accents}

\usepackage{tikz}
\usetikzlibrary{automata,topaths}
\usetikzlibrary{shapes}
\usetikzlibrary{plotmarks}
 
\usepackage{float} 
\usepackage{textcomp} 
\usepackage{siunitx}
\usepackage{xcolor}

\definecolor{lightblue}{rgb}{0,0.2,0.5}
\usepackage[colorlinks=true, urlcolor=lightblue,linkcolor=lightblue, citecolor=lightblue]{hyperref}
\usepackage{ucs}

\definecolor{dblackcolor}{rgb}{0.0,0.0,0.0}
\definecolor{dbluecolor}{rgb}{0.01,0.02,0.7}
\definecolor{dgreencolor}{rgb}{0.2,0.4,0.0}
\definecolor{dgraycolor}{rgb}{0.30,0.3,0.30}

\definecolor{ForestGreen}{RGB}{34,139,34}
\definecolor{mauve}{rgb}{0.7,0,0.43}
\definecolor{dkgreen}{rgb}{0,0.6,0}
\definecolor{darkgreen}{rgb}{0,0.6,0}
\definecolor{darkorange}{rgb}{1.0, 0.55, 0.0}
\definecolor{lightblue}{rgb}{0,0.2,0.5}
\definecolor{blue1}{rgb}{0,0.1,0.9}

\definecolor{lightblue}{rgb}{0,0.2,0.5}

\usepackage{listings}
\lstdefinelanguage{Maple}{
    morekeywords={proc, if, return, map, op, int, for, do, local, nops, convert, end},
    sensitive=false, %
    morecomment=[l]{//}, %
    morecomment=[s]{/*}{*/}, %
    morestring=[b]" %
} %

\usepackage{graphicx}
\usepackage{flushend,cuted}
\usepackage{bm}

\usepackage{tabularx}
\usepackage{indentfirst}
\usepackage{amssymb}
\usepackage{xparse}
\usepackage{tikz}
\usepackage{mdwlist}
\usepackage{tkz-graph}

\DeclareMathAlphabet{\eufrak}{U}{}{}{} 
\SetMathAlphabet\eufrak{normal}{U}{euf}{m}{n}
\SetMathAlphabet\eufrak{bold}{U}{euf}{b}{n}

\newcommand{\E}{\mathbb{E}}

\newcommand{\N}{\mathbb{N}}
\newcommand{\inte}{\mathbb{N}}

\makeatletter
\newcommand*\rel@kern[1]{\kern#1\dimexpr\macc@kerna}
\newcommand*\widebar[1]{%
  \begingroup
  \def\mathaccent##1##2{%
    \rel@kern{0.8}%
    \overline{\rel@kern{-0.8}\macc@nucleus\rel@kern{0.2}}%
    \rel@kern{-0.2}%
  }%
  \macc@depth\@ne
  \let\math@bgroup\@empty \let\math@egroup\macc@set@skewchar
  \mathsurround\z@ \frozen@everymath{\mathgroup\macc@group\relax}%
  \macc@set@skewchar\relax
  \let\mathaccentV\macc@nested@a
  \macc@nested@a\relax111{#1}%
  \endgroup
}
\makeatother

\makeatletter
\DeclareRobustCommand\widecheck[1]{{\mathpalette\@widecheck{#1}}}
\def\@widecheck#1#2{%
    \setbox\z@\hbox{\m@th$#1#2$}%
    \setbox\tw@\hbox{\m@th$#1%
       \widehat{%
          \vrule\@width\z@\@height\ht\z@
          \vrule\@height\z@\@width\wd\z@}$}%
    \dp\tw@-\ht\z@
    \@tempdima\ht\z@ \advance\@tempdima2\ht\tw@ \divide\@tempdima\thr@@
    \setbox\tw@\hbox{%
       \raise\@tempdima\hbox{\scalebox{1}[-1]{\lower\@tempdima\box
\tw@}}}%
    {\ooalign{\box\tw@ \cr \box\z@}}}
\makeatother

\newtheorem{prop}{Proposition}[section]
\newtheorem{lemma}[prop]{Lemma}

\newtheorem{corollary}[prop]{Corollary}
\newtheorem{theorem}[prop]{Theorem}

\usepackage[title]{appendix} %

\newenvironment{Proof}{\removelastskip\par\medskip
\noindent{\em Proof.} \rm}{\penalty-20\null\hfill$\square$\par\medbreak}

\newenvironment{Proofy}{\removelastskip\par\medskip
\noindent{\em Proof\hskip-0.12cm} \rm}{\penalty-20\null\hfill$\square$\par\medbreak}

\def\({\left(}
\def\){\right)}

\def\[{\left[}
\def\]{\right]}
\def\real{{\mathord{\mathbb R}}}
\def\N{{\mathord{\mathbb N}}}

\def\P{\mathbb{P}}

\makeatletter
\newcommand*\bigcdot{\mathpalette\bigcdot@{.5}}
\newcommand*\bigcdot@[2]{\mathbin{\vcenter{\hbox{\scalebox{#2}{$\m@th#1\bullet$}}}}} %
\makeatother

\usepackage{cprotect}

\usepackage{empheq}

\usepackage{scalerel} 

\usepackage{comment}

\newcommand{\ind}{\mathbf 1}

\newcommand{\pt}{\partial}

\allowdisplaybreaks

\numberwithin{equation}{section}

\GraphInit[vstyle = Shade]
\usetikzlibrary[intersections,
positioning,
petri,
backgrounds,
fit,
decorations.pathmorphing,
arrows,
arrows.meta,
bending,
calc,
intersections,
through,
backgrounds,
shapes.geometric,
quotes,
matrix,
trees,
shapes.symbols,
graphs,
math,
patterns,
external,
scopes,
matrix,
lindenmayersystems,
shapes.callouts,
shapes.misc,
angles,
shapes.arrows,
shadings]

\begin{document}

\title{
\huge
Binary Galton--Watson trees with mutations 
} 

\author{
  Qiao Huang\footnote{
  School of Mathematics,
  Southeast University,
  Nanjing 211189,
  P.R. China,
  and 
 School of Physical and Mathematical Sciences, 
 Nanyang Technological University, 
 21 Nanyang Link, Singapore 637371. 
 \href{mailto:qiao.huang@seu.edu.cn}{qiao.huang@seu.edu.cn}}
 \qquad
 Nicolas Privault\footnote{%
 School of Physical and Mathematical Sciences, 
 Nanyang Technological University, 
 21 Nanyang Link, Singapore 637371.
 \href{mailto:nprivault@ntu.edu.sg}{nprivault@ntu.edu.sg}
}
}

\maketitle

\vspace{-0.5cm}

\begin{abstract} 
 We consider a multitype Galton--Watson process 
 {that} allows for the mutation and reversion
 of individual types in discrete and continuous time.
 In this setting, we explicitly compute 
 the time evolution of quantities such as the
 mean and distributions of different types. 
 This allows us in particular to estimate the proportions
 of different types in the long run,
 as well as the distribution of the first time of occurrence of
 a given type as the tree size or time increases.
 Our approach relies on the recursive computation of 
 the joint distribution of types conditionally to the value of
 the total progeny.
 In comparison with the literature on related multitype models,
 we do not rely on approximations. 
\end{abstract}


\medskip

\baselineskip0.7cm

\section{Introduction} 
\noindent
  Multitype Galton--Watson processes
  have been used in population genetics and
  evolutionary biology
  to model the propagation and extinction of 
  mutant types.
  In \cite{iwasa} and \cite{durrett2},
  the mutation of a type-$i$ cancer cell mutates into a type $i + 1$
  cell has been modeled using a continuous-time process
  that branches at an exponential rate depending on $i\geq 0$.
 In \cite{sagitov}, evolutionary branching processes
 modeling subpopulations with
 different traits or genotypes  
 have been analyzed under small mutational step sizes.
 More recently, the diffusion limit of 
 Galton--Watson branching processes 
 modeling alele types has been analyzed in 
 \cite{cjburden}. 
 This analysis relies mainly 
 on the classical literature on birth-death
 and Galton--Watson processes, see e.g. 
 \cite{kendall1948},
 \cite{Ott49},
 \cite{harris1963}, 
 \cite{athreya}, and 
 \cite{brown-shubert}, 
 which is used to derive the distribution and generating function
 properties of the progeny of random trees or branching
 processes. %

\medskip

 In this paper, we consider multitype random binary trees
 in which every node
 bears an integer type $j\geq 0$ and may generate
 either no offspring, or two offsprings, 
 one with type $0$ and one with type $j+1$. %
 Our analysis covers both the discrete-time case,
 in which two offsprings are generated with probability $p\in (0,1)$
 and no offspring is generated with probability $q:=1-p$,
 and the continuous-time setting, which yields
 a multitype Galton---Watson {process}. 

 \medskip 

Figure~\ref{fig1} presents a sample of the corresponding 
 multitype branching process 
 in discrete time. %

\begin{figure}[H]
\tikzstyle{level 1}=[level distance=6cm, sibling distance=4cm]
\tikzstyle{level 2}=[level distance=6cm, sibling distance=3.5cm]
\tikzstyle{level 3}=[level distance=6cm, sibling distance=2cm]
\tikzstyle{level 4}=[level distance=6cm, sibling distance=2cm]
\tikzstyle{level 5}=[level distance=6cm, sibling distance=2cm]
\begin{center}
\resizebox{0.6\textwidth}{!}{
\begin{tikzpicture}[scale=1.0,grow=right, sloped]
                \node[scale=2,rectangle,draw,black,text=black,very thick, fill=green!60] {$3$} %
                child{
                node[scale=2,rectangle,draw,black,text=black,thick,yshift=-0.4cm, fill=orange!60]{$4$} %
                    child{
                    node[scale=2,rectangle,draw,black,text=black,thick, fill=purple!60]{$5$} %
                    child{
                    node[scale=2,rectangle,draw,black,text=black,thick, fill=black!50]{$6$} %
                    edge from parent
                    node[above]{$ $} %
                    node[below]{$ $}
                    }
                    child{
                    node[scale=2,rectangle,draw,black,text=black,thick,fill=gray!15]{$0$}
                    child{
                    node[scale=2,rectangle,draw,black,text=black,thick, fill=pink]{$1$}
                    edge from parent
                    node[above]{$ $} %
                    node[below]{$ $}
                    }
                    child{
                    node[scale=2,rectangle,draw,black,text=black,thick,fill=gray!15]{$0$}
                    edge from parent
                    node[above]{$ $} %
                    node[below]{$ $}
                    }
                    edge from parent
                    node[above]{$ $} %
                    node[below]{$ $}
                    }
                    edge from parent
                    node[above]{$ $} %
                    node[below]{$ $}
                    }
                    child{
                    node[scale=2,rectangle,draw,black,text=black,thick,fill=gray!15]{$0$}
                    edge from parent
                    node[above]{$ $} %
                    node[below]{$ $}
                    }
                edge from parent
                node[above]{$ $} %
                node[below]{$ $}
                }
                child{
                node[scale=2,rectangle,draw,black,text=black,thick,yshift=0.4cm,fill=gray!15]{$0$}
                    child{
                    node[scale=2,rectangle,draw,black,text=black,thick, fill=pink]{$1$}
                    child{
                    node[scale=2,rectangle,draw,black,text=black,thick, fill=blue!40]{$2$}
                    child{
                    node[scale=2,rectangle,draw,black,text=black,thick, fill=green!60]{$3$}
                    child{
                    node[scale=2,rectangle,draw,black,text=black,thick, fill=orange!60]{$4$}
                    edge from parent
                    node[above]{$ $} %
                    node[below]{$ $}
                    }
                    child{
                    node[scale=2,rectangle,draw,black,text=black,thick,fill=gray!15]{$0$}
                    edge from parent
                    node[above]{$ $} %
                    node[below]{$ $}
                    }
                    edge from parent
                    node[above]{$ $} %
                    node[below]{$ $}
                    }
                    child{
                    node[scale=2,rectangle,draw,black,text=black,thick,fill=gray!15]{$0$}
                    edge from parent
                    node[above]{$ $} %
                    node[below]{$ $}
                    }
                    edge from parent
                    node[above]{$ $} %
                    node[below]{$ $}
                    }
                    child{
                    node[scale=2,rectangle,draw,black,text=black,thick,fill=gray!15]{$0$}
                    edge from parent
                    node[above]{$ $} %
                    node[below]{$ $}
                    }
                    edge from parent
                    node[above]{$ $} %
                    node[below]{$ $}
                    }
                    child{
                    node[scale=2,rectangle,draw,black,text=black,thick,fill=gray!15]{$0$}
                    child{
                    node[scale=2,rectangle,draw,black,text=black,thick, fill=pink]{$1$}
                    edge from parent
                    node[above]{$ $} %
                    node[below]{$ $}
                    }
                    child{
                    node[scale=2,rectangle,draw,black,text=black,thick,fill=gray!15]{$0$}
                    edge from parent
                    node[above]{$ $} %
                    node[below]{$ $}
                    }
                    edge from parent
                    node[above]{$ $} %
                    node[below]{$ $}
                    }
                edge from parent
                node[above]{$ $} %
                node[below]{$ $}
            };
\end{tikzpicture}
}
\end{center}
\vspace{-0.2cm}
\caption{Marked random tree sample started from the initial type $j=3$.
}
\label{fig1}
\end{figure}

\vspace{-0.3cm}

 In terms of population genetics,
 such trees provide a way to model mutation and reversion,
 by considering ``wild type'' individuals with type $0$, and ``mutant''
 individuals with type $j\geq 1$
 which have been the object of $j\geq 1$
 mutations. 
 In this setting, 
 wild type $0$ individuals can have offspring of both
 wild type $0$ and mutant type $1$,
 whereas mutants of type $j\geq 1$ can 
 have offsprings of wild type $0$ (revertants), or mutant
 type $j+1$. See for example
 \cite{avezedo} 
 and references therein for the study of
 related population models in the framework
 of evolutionary rescue. 

 \medskip

 In this model,
   type $0$ individuals follow the same Galton--Watson
   dynamics
 as in \cite{iwasa}, \cite{durrett2},
 and the interbranching times are also exponentially
   distributed.
   However, higher type transitions are modeled
   differently using a marked tree,
   and the growth of population types over times
   can be subexponential, see Figure~\ref{fig2-11-2-0},
   whereas it is exponential in
   Theorem~4 of \cite{durrett2}.
   In addition, due to model complexity,
   the computation of waiting times distributions in 
   \cite{iwasa}, \cite{durrett2} {relies}
   on approximation formulas.
   
  \medskip

 Our main results 
 are presented in Sections~\ref{s2} and \ref{s3} 
 and are proved in 
 Appendices~\ref{s4} and \ref{s5}, respectively 
 in discrete and continuous time. 
 After recalling the computation of the distribution of
 tree progenies in Propositions~\ref{radius-conv-2}
 and \ref{progeny}, 
 we derive recursive expressions for %
 the distribution of any finite vector
 $\big(X^{(1)},\ldots , X^{(n)}\big)$ of type counts given
 the size of the random tree, %
 see Theorems~\ref{const-weight-dc} and \ref{const-l-weight}, 
 respectively in discrete and continuous time. 
 
 \medskip

 In particular, we provide closed form and
   recursive expressions for 
   the distribution of type counts in both
   discrete and continuous time,
   which allows us to determine the evolution of quantities such
   as:
   \begin{itemize}
     \item expected counts and proportions 
   of types, see 
   Figures~\ref{fig2}, \ref{fig2-2},
   {\ref{fig2-11-2-0} and \ref{fig2-11-2}}, 
   and
   \item the distributions of the first 
   occurrence times of a given type count,
   see {Figures~\ref{fig2-11-2-0-2}
     and \ref{fig2-11-2-0-3}}. 
   \end{itemize}
    We also derive identities for
 the expectation of product functionals on random trees,
 which in turn yield integrability conditions
 for generating functionals,
 see Sections~\ref{s3-1} and \ref{s3-3}. 
 
\medskip
  
 In Figure~\ref{fig2} we display the computed values of
 the conditional mean proportions of types as the
 size of the discrete-time tree increases.
 Figure~\ref{fig2-11-2} displays the continuous-time evolution of those
 {proportions}. 
 We note in particular
 that the (wild) type $0$ remains predominant in discrete time, see 
 Figure~\ref{fig2}, whereas in continuous time 
 it is the initial type $j$ which remains predominant 
 over time in Figure~\ref{fig2-11-2}. 
 Figures~\ref{fig2-11-2-0-2} and \ref{fig2-11-2-0-3} 
 present the tail distribution functions
 and probability density functions of the occurrence times
 of given types. 

 \medskip

 The closed-form expressions
 of Theorems~\ref{const-weight-dc} and \ref{const-l-weight} 
 are then applied to the computation of
 the expectation of product functionals on random trees
 in Proposition~\ref{B-special} 
 and Corollaries~\ref{c1}, \ref{c2}
 in discrete time, and in 
 Propositions~\ref{hjkfd21}, \ref{A-special}
 and Corollary~\ref{c3} in continuous time.
 In particular, Corollaries~\ref{c2} 
 and \ref{c3} yield sufficient conditions 
 for the integrability of random
 product functionals involving marks.
 Such results are applicable to problems 
 where the generation of random trees
 is used in Monte Carlo integration,
 see for example \cite{huangprivault}, 
 \cite{huangprivault2}
 for an application to Monte Carlo methods
 for differential equations.

\medskip

 The recurrence formulas proved in Theorems~\ref{const-weight-dc}
 and \ref{const-l-weight} 
 are implemented in Mathematica {and Python} notebooks
 which are used to produce
 Figures~\ref{fig1}-\ref{fig2-2-1}  
 and \ref{fig2-11-2-0}-\ref{fig2-11-2}, and  
 are available at %

\centerline{\href{https://github.com/nprivaul/branching/}{https://github.com/nprivaul/branching/}}

\noindent in discrete and continuous time.  %

 \medskip

 \noindent
 All analytical results are 
 confirmed by Monte Carlo simulations that can
 also be run in the above notebooks.
 
\section{Discrete-time setting}
\label{s2}
\noindent
{In what follows, we let $\inte = \{0,1,2,\ldots \}$}.
\subsection{Marked Galton--Watson process}  
\noindent
 We consider a branching chain
 in which every individual has either no offspring with probability $q$,
 or two offsprings with probability $p$.
 For this, let $({\xi}_{n,k})_{n, k\geq 1}$ denote a family
 of independent $\{0,2\}$-valued Bernoulli random variables with
 the common distribution
$$
q= \P({\xi}_{n,k} = 0) \quad
\mbox{and}
\quad 
p= \P({\xi}_{n,k} = 2), \quad n,k \geq 1, 
$$
with $p+q=1$ and $0<p,q<1$,
where $\xi_{n,k}$ represents
the number of offsprings of the $k$-$th$ individual
 of generation $n-1$, %
 see e.g. \cite{harris1963}, \cite{athreya}.

\medskip
 
In this framework, the branching chain $({Z}_n)_{n\geq 0}$ is recursively defined as 
\begin{equation}
Z_0 = 1, \quad Z_n = \sum_{k=1}^{Z_{n-1}} {\xi}_{n,k}, \quad n\geq 1, 
\end{equation}
and represents the population size at generation $n\geq 0$.
 We let
$$
S^{\scaleto{\neq 0}{6pt}}_\infty := \frac{1}{2} \sum_{k=1}^\infty {Z}_k
$$
 denote the total count of nodes with non-zero types,
 excluding the initial node,
 i.e. $1+2 S^{\scaleto{\neq 0}{6pt}}_\infty$ represents the total progeny of the chain
 $({Z}_n)_{n\geq 0}$.
 Note that $S^{\scaleto{\neq 0}{6pt}}_\infty$ 
 is also equal to the number of nodes with type $0$
 excluding the initial node, since each node with non-zero
 type has a co-twin with type $0$.

\medskip

Using the sequence $({\xi}_{n,k})_{n,k\geq 1}$
we construct a marked random binary tree
${\cal T}$ in which a node
 $k\in \{1,\ldots , Z_{n-1}\}$ at generation $n-1$
yields either two branches if $\xi_{n,k}=2$, or 
zero branch if $\xi_{n,k}=0$. 
In addition, the nodes of ${\cal T}$ receive marks 
that represent individual types, as described below. 
\begin{enumerate}[i)]
 \item
   The initial node %
   has the type 
   $j\geq 0$;
 \item
   if a node %
   of type %
   $i\in\N$ splits,
   its two offsprings respectively
   receive the types $0$ and $i+1$; 
 \end{enumerate}
 as shown in Figure~\ref{fig1}.
\noindent
 Proposition~\ref{radius-conv-2} recovers the distribution
 of the number of vertices of the random binary tree
 ${\cal T}$ using classical results of \cite{Ott49},
 and is proved in Appendix~\ref{s4} for completeness. 
 In what follows, we let 
 $$
 C_n 
  = \frac{1}{n+1} \binom{2n}{n} 
 , \quad n \geq 0, 
 $$ 
 denote the $n$-$th$ Catalan number (see \cite{Aig07}),
 which represents the number of different rooted binary trees
 with $n+1$ leaves. %
\begin{prop}
\label{radius-conv-2}
 The distribution of
 the count $S^{\scaleto{\neq 0}{6pt}}_\infty$
 of nodes with non-zero types 
 is given by 
  \begin{equation}
    \label{recursion-total-sol-discrete}
 \P\big( S^{\scaleto{\neq 0}{6pt}}_\infty=n\big) = 
q (pq)^n C_n , \quad  n\geq 0,
\end{equation}
 with the probability generating function 
 \begin{equation}
   \label{fjkldf34} 
   \E \big[ \delta^{S^{\scaleto{\neq 0}{4.5pt}}_\infty} \big]
   = \frac{1-\sqrt{1-4 pq \delta }}{2 p \delta },
   \quad |\delta | \leq 1/(4pq), 
\end{equation}
 and we have $\P(S^{\scaleto{\neq 0}{6pt}}_\infty < \infty ) = 1$
 if $p\leq 1/2$. 
\end{prop}
\noindent
 In addition, it follows from \eqref{fjkldf34} that if $p<1/2$, we have 
 \begin{equation}
   \label{fjkl242}
   \E[S^{\scaleto{\neq 0}{6pt}}_\infty ] = p/(q-p). 
   \end{equation} 
\subsection{Conditional type distribution} 
\noindent
 We let $X^{(k)}$ denote the count of types equal to $k \geq 1$ %
 in the random tree ${\cal T}$, excluding the initial node, with 
$$
 X^{(k)} = 0 \mbox{ for } k > S^{\scaleto{\neq 0}{6pt}}_\infty .
$$ 
 For example, in Figure~\ref{fig1}
 with $j = 3$ we have $S^{\scaleto{\neq 0}{6pt}}_\infty = 9$, and 
$$
 X^{(1)}=3, \ X^{(2)}=1, \ X^{(3)}=1, \ X^{(4)}=2, \ X^{(5)}=X^{(6)}=1. 
$$ 
 We also let $\P_j$, resp. $\E_j$,
 denote conditional probabilities
 and expectations given that ${\cal T}$
 is started from the initial type $j\in \inte$. 

 \medskip
 
\begin{figure}[H]
  \centering
 \begin{subfigure}[b]{0.45\textwidth}
    \includegraphics[width=1\linewidth, height=3.9cm]{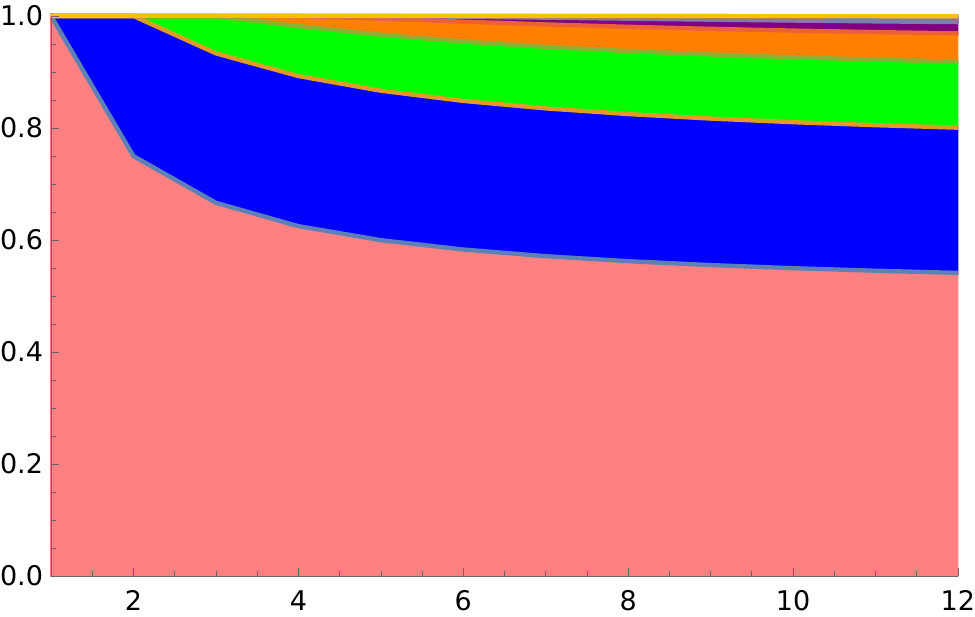}
    \caption{Initial type $j=0$.} 
 \end{subfigure}
 \begin{subfigure}[b]{0.45\textwidth}
    \includegraphics[width=1\linewidth, height=3.9cm]{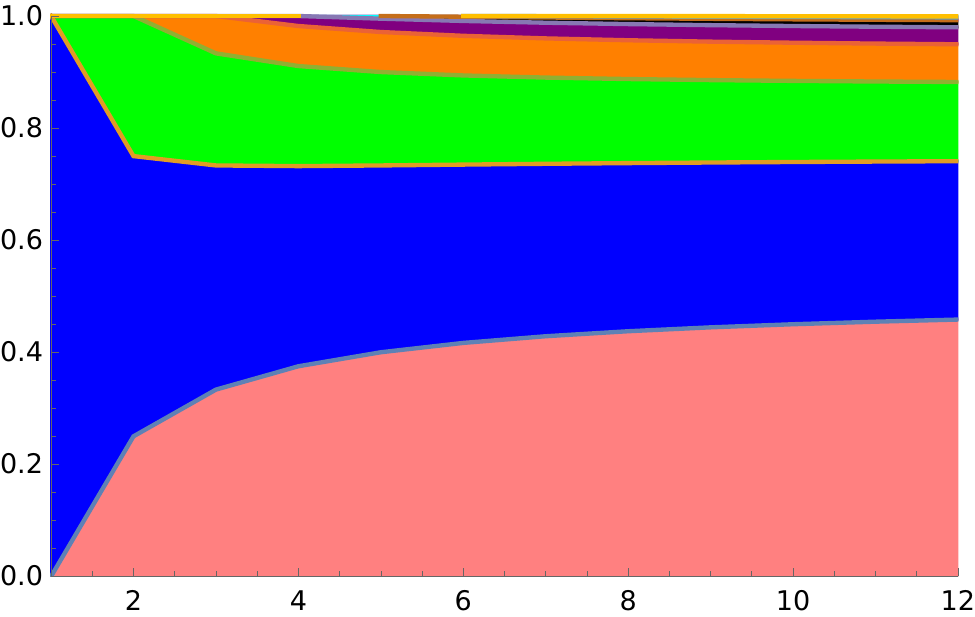}
    \caption{Initial type $j=1$.} 
 \end{subfigure}
 \begin{subfigure}[b]{0.45\textwidth}
    \includegraphics[width=1.0\linewidth, height=3.9cm]{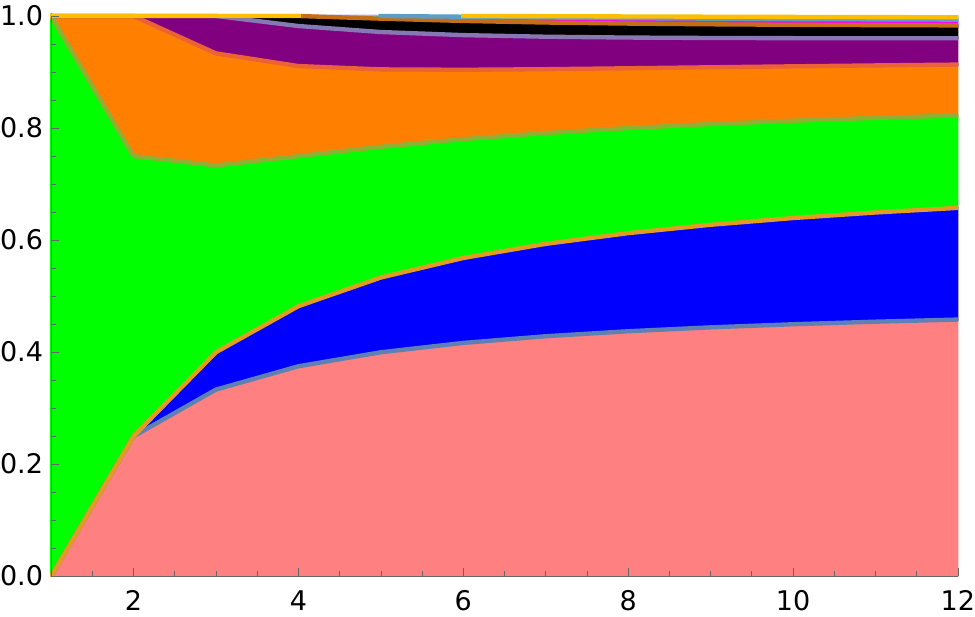}
    \caption{Initial type $j=2$.} 
 \end{subfigure}
  \centering
  \begin{subfigure}[b]{0.45\textwidth}
 \hskip0.1cm
    \includegraphics[width=1.0\linewidth, height=3.9cm]{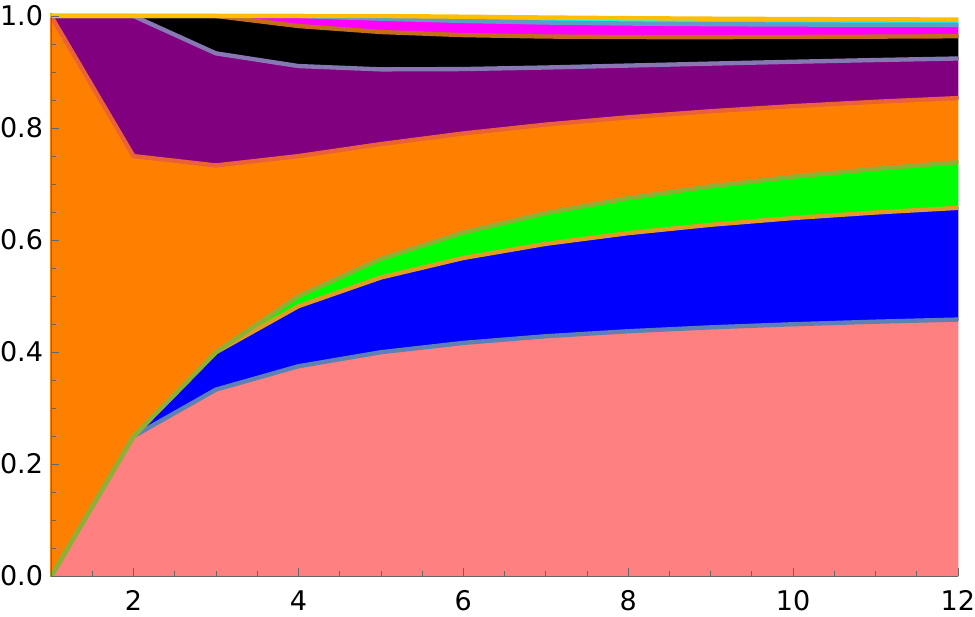}
    \caption{Initial type $j=3$.} 
 \end{subfigure}
  \caption{Conditional average type proportions
    \eqref{fjkld3-1} given
 the values of $S^{\scaleto{\neq 0}{6pt}}_\infty$ in abscissa. %
} 
\label{fig2} 
\end{figure}

 In Theorem~\ref{const-weight-dc},
 which is proved in Appendix~\ref{s4},
 we compute recursively
 the conditional type distribution
 of $\big( X^{(1)} ,\ldots , X^{(n)}\big)$ given
 that their summation
 equals $S^{\scaleto{\neq 0}{6pt}}_\infty$
 and $X^{(k)}=0$ for all $k>n$, and show that it does
 not depend on $p,q$.
 In what follows, we use the notation
 $\ind_A$ to denote the indicator function
 taking the value $1$, resp. $0$ when condition~A
 is satisfied, resp. not satisfied. 
\begin{theorem}
\label{const-weight-dc}
For $j \geq 0$ {and $n\geq 1$}, the conditional type distribution 
 \begin{equation}
 \label{recursion-1-sol-j-0}
  \P_j \big( X^{(1)} = m_1,\ldots , X^{(n)}=m_n
  \mid S^{\scaleto{\neq 0}{6pt}}_\infty = m_1+\cdots + m_n \big) 
  =
  \frac{b_j(m_1,\ldots,m_n)}{C_{m_1+\cdots + m_n}},   
\end{equation}
 $m_1,\ldots , m_n\geq 0$, %
 {satisfies} the {recursive relation} 
 \begin{equation}
   \label{recursion-1-coef-general-discrete-j}
  \hskip-0.3cm
  b_j(m_1,\ldots,m_n) = \sum_{l=1}^{n-j}
  \ind_{\{m_{j+l} > m_{j+l+1}\}} %
  \sum_{\begin{subarray}{c}
      \sum_{k=1}^l m_i^k = m_i -\ind_{\{j < i\leq j+l\}}, \ \! 1\leq i\leq n \\
      0\leq m_i^k \leq m_{i-1}^k, \ \! 2\leq i\leq n, \ \! 1\leq k\leq l
    \end{subarray}} \prod_{k=1}^l b_0\big(m^k_1,\ldots,m^k_n\big)
\end{equation}
where $m_{n+1}:=0$
{in the last indicator  
 $\ind_{\{m_n > m_{n+1}\}}$}, 
$$
b_j(\varnothing) = 1,
\quad 
b_j(m_1,\ldots,m_{n-1},0)
= b_j(m_1,\ldots,m_{n-1}),
$$ 
 and
 $$
 b_j(m_1,\ldots,m_n) = 0 \mbox{ if } 1\leq n < j
 \mbox{ and }
 {m_1 + \cdots + m_n \geq 1}.
$$ 
\end{theorem}
\noindent
{From Theorem~\ref{const-weight-dc},
 the conditional type distribution \eqref{recursion-1-sol-j-0} 
 can be computed first by applying the recursion 
 \eqref{recursion-1-coef-general-discrete-j}
 to $j=0$, as
$$
 \hskip-0.3cm
  b_0(m_1,\ldots,m_n) = \sum_{l=1}^n \ind_{\{m_l > m_{l+1}\}} %
  \sum_{\begin{subarray}{c}
      \sum_{k=1}^l m_i^k = m_i -\ind_{\{1 \leq i\leq l\}}, \ \! 1\leq i\leq n \\
      0\leq m_i^k \leq m_{i-1}^k, \ \! 2\leq i\leq n, \ \! 1\leq k\leq l
    \end{subarray}} \prod_{k=1}^l b_0\big(m^k_1,\ldots,m^k_n\big), 
$$ 
 and then applying again 
 \eqref{recursion-1-coef-general-discrete-j}
 to $j\geq 1$. 
 Note also that when $m_1+\cdots + m_n < n - j$,
 the summation range in
 \eqref{recursion-1-coef-general-discrete-j}
 is empty,
 whence
 $b_j(m_1,\ldots , m_n)=0$. 
 In addition,} setting 
 $$
 {\mathbb K}_{j,n} := 
 \left\{
 \begin{array}{ll} 
 \{\varnothing\}\cup \{(m_1,\ldots,m_n)
  \ \! : \ \! 
  m_1\ge \cdots\ge m_n \ge 1 \}, 
 & 
 j=0, \ n\geq 0, 
  \medskip
\\
 \{(m_1,\ldots,m_n) \ \! : \ \! 
  m_1\geq \cdots\geq m_j \geq 0, \
  m_j +1 \geq m_{j+1}\ge \cdots \ge m_n\ge 1 \},
  &
  1 \leq j < n,
  \medskip
\\ 
\{(m_1,\ldots,m_j) \ \! : \ \! m_1= \cdots = m_j = 0 \},
& j=n\geq 1, 
  \medskip
  \\
  \emptyset,
    & 1 \leq n < j,
 \end{array}
 \right.
 $$ 
  for $j\geq 0$, $1 \leq n \leq m+j$, 
  {$m\geq 1$},
  and any weight function $f_n : \inte^n \to \real$,
  we have 
\begin{equation*}
 \E_j \big[
   f_n \big( X^{(1)},\ldots , X^{(n)} \big)
   {\bf 1}_{\{
     X^{(1)} + \cdots + X^{(n)} = m %
     \}}      
   \ \! \big| \ \!
   S^{\scaleto{\neq 0}{6pt}}_\infty = m
   \big]
 =
 \hskip-0.5cm
 \sum_{\begin{subarray}{c}
    (m_1,\ldots,m_n)\in {\mathbb K}_{j,n} \\
    m_1 + \cdots + m_n = m
    \end{subarray}}
 \hskip-0.3cm
  \frac{b_j(m_1,\ldots,m_n)}{C_m}
 f_n (m_1,\ldots , m_n). 
\end{equation*}
 In particular, the following corollary
 provides a way to solve the recursion
 \eqref{recursion-1-coef-general-discrete-j}
 for the computation of mean type counts given
 the value of $S^{\scaleto{\neq 0}{6pt}}_\infty$. 
\begin{corollary} 
\label{fjklds13-0}
 The mean count of type $l$
 individuals given the value of $S^{\scaleto{\neq 0}{6pt}}_\infty = m$
 after starting from type $j$ is given by 
\begin{align} 
\label{fjkld3-1}
 \E_j \big[ X^{(l)} \ \! \big| \ \! S^{\scaleto{\neq 0}{6pt}}_\infty = m
   \big]
 =
 \frac{1}{C_m}
 \ind_{\{ 0 < l-j \leq m \}}
          \frac{l-j+1}{m+1}
          {2m-l+j \choose m}
          +
          \frac{1}{C_m}
          \ind_{\{ m \geq l \}}
          {2m-l \choose m+1}, 
\end{align} 
$l,m\geq 1$, $j\geq 0$.
\end{corollary}
\noindent 
 The proof of Corollary~\ref{fjklds13-0},
 which is given in Appendix~\ref{s4},
 also shows that
$$ 
 \E_j \big[ X^{(l)} \big]
 =
 \ind_{\{ j < l\}}
  p^{l-j}
  +
  \frac{p^{l+1}}{q-p},
  \quad j\geq 0, \ l\geq 1,
  $$
  and
  $$
  \sum_{k\geq 1}
  \E_j \big[ X^{(k)} \big]
  =
  \frac{p}{q-p},
$$
 which recovers \eqref{fjkl242}, with
 $$ 
 \frac{\E_j \big[ X^{(l)} \big]}{
  \sum_{k\geq 1} \E_j \big[ X^{(k)} \big]}
  =
 \ind_{\{ j < l\}}
  (q-p) p^{l-j-1}
  +
  p^l,
      \quad j\geq 0, \ l\geq 1.
$$ 
 As a consequence of Corollary~\ref{fjklds13-0},
 the conditional mean proportions of non-zero types 
\begin{equation}
\label{fjkl1243} 
 \frac{1}{m}
 \E_j \big[ X^{(l)} \ \! \big| \ \! S^{\scaleto{\neq 0}{6pt}}_\infty = m
   \big], \qquad m\geq 1, 
\end{equation} 
 satisfy 
\begin{equation*}
  \lim_{m\to \infty}
 \frac{1}{m}
 \E_j \big[ X^{(l)} \ \! \big| \ \! S^{\scaleto{\neq 0}{6pt}}_\infty = m
   \big]
 = \frac{1}{2^l}, \quad j\geq 0, \ l\geq 1. 
\end{equation*}
\noindent 
\noindent 
Figure~\ref{fig2} displays the evolution of computed values of 
 the conditional mean proportions
 \eqref{fjkl1243} of non-zero types
 for $m=1,\ldots , 12$, 
 after starting from the initial types $j=0,1,2,3$. %

\vspace{-0.3cm}

\noindent
 The color coding of types used in Figures~\ref{fig1}-\ref{fig2-2}
 and \ref{fig2-11-2-0}-\ref{fig2-11-2} is shown below. 

\vspace{-0.6cm}

\begin{equation}
    \label{dkjl}
    \begin{tabular}{|c|c|c|c|c|c|c|c|c|} %
\hline
 \cellcolor{black}  \textcolor{white}{0} \! \! &  \cellcolor{pink!70} 1 \! \! & \cellcolor{blue!60} 2 \! \! & \cellcolor{green!60} 3 \! \! & \cellcolor{orange!60} 4 \! \! & \cellcolor{purple!70} 5 \! \! & \cellcolor{black!50} 6 \! \! & \cellcolor{magenta!50} 7 \! \! & \cellcolor{blue!40} 8 \! \! 
\\
\hline
\end{tabular}
\end{equation}

\noindent
 The expected values of the conditional proportions
 \eqref{fjkl1243} of non-zero types are computed
 as functions of $p \in (0,1/2)$
 in Corollary~\ref{fjklds13-1}.
 Here, 
 $$
 {\rm B} (z;a,b): =
 \int_0^z u^{a-1} (1-u)^{b-1}
 du
 $$
 denotes the incomplete beta function. 
\begin{corollary} 
\label{fjklds13-1}
 The conditional proportion of type $k$
 individuals after starting from the initial type $j$ is given by 
\begin{equation} 
\label{fjkld3-2}
   \E_j \left[ \frac{X^{(k)}}{
        S^{\scaleto{\neq 0}{6pt}}_\infty }
      \ \! \Big| \ \!
      S^{\scaleto{\neq 0}{6pt}}_\infty \geq 1
      \right] 
   =
     \frac{q}{p}
{\rm B}( p ; k+1, -1 ) 
   +
   \frac{q}{p}
   \ind_{\{k>j \}}
   \left(
  (k+1-j) 
  {\rm B} ( p ; k-j, 0 )
   - \frac{p^{k-j}}{q}
   \right)
, 
\end{equation} 
 $k\geq 1$, $j\geq 0$. 
\end{corollary}
\noindent
The proof of Corollary~\ref{fjklds13-1} is
 {given} in Appendix~\ref{s4},
 and the average proportions \eqref{fjkld3-2}  
 are plotted in Figure~\ref{fig2-2} for
 $p\in [0,1/2)$, after starting from
 the initial types $j=0,1,2,3$. %

 \begin{figure}[H]
  \centering
 \begin{subfigure}[b]{0.45\textwidth}
    \includegraphics[width=1\linewidth, height=3.9cm]{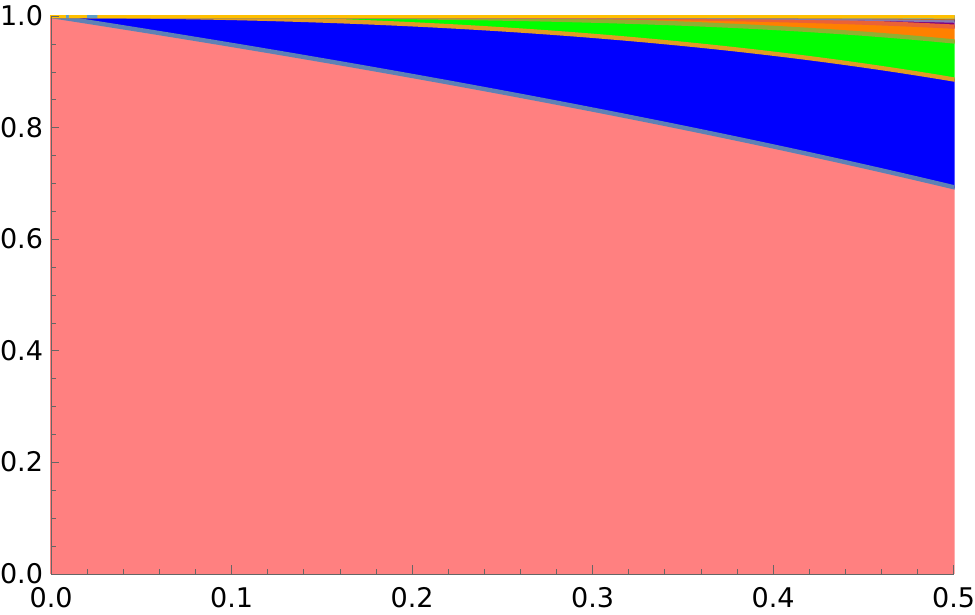}
    \caption{Initial type $j=0$.} 
 \end{subfigure}
 \begin{subfigure}[b]{0.45\textwidth}
    \includegraphics[width=1\linewidth, height=3.9cm]{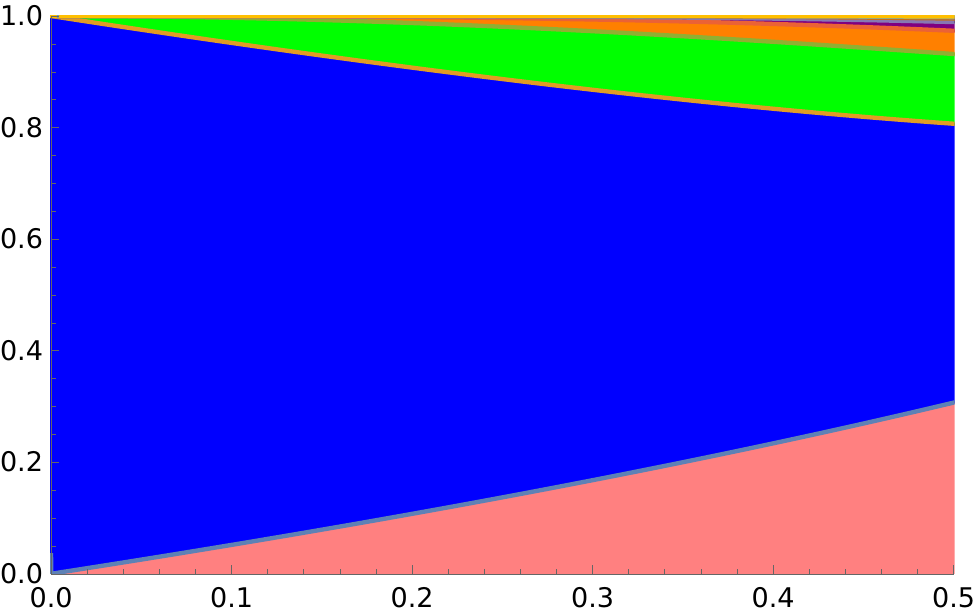}
    \caption{Initial type $j=1$.} 
 \end{subfigure}
 \begin{subfigure}[b]{0.45\textwidth}
    \includegraphics[width=1.0\linewidth, height=3.9cm]{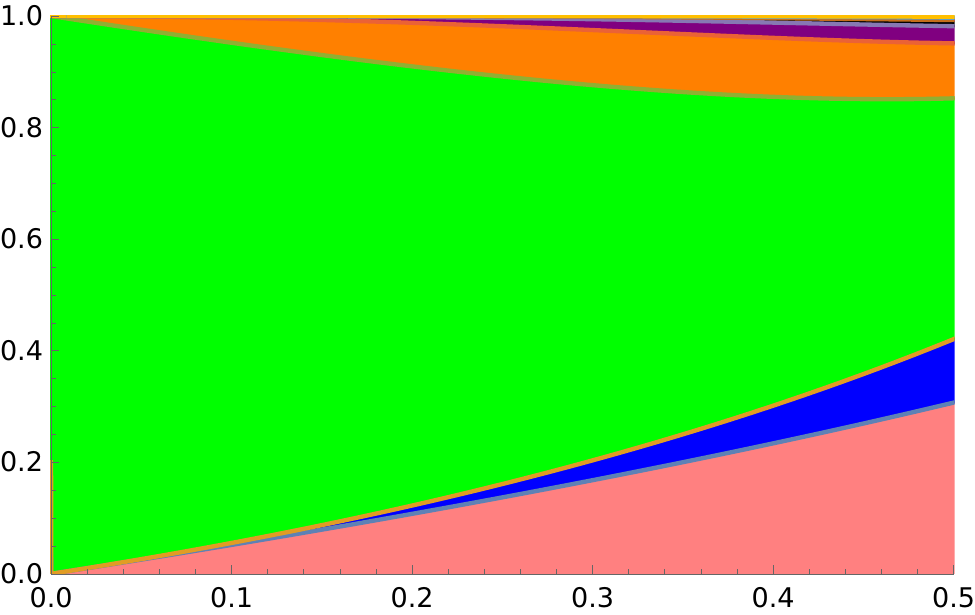}
    \caption{Initial type $j=2$.} 
 \end{subfigure}
  \centering
  \begin{subfigure}[b]{0.45\textwidth}
 \hskip0.1cm
    \includegraphics[width=1.0\linewidth, height=3.9cm]{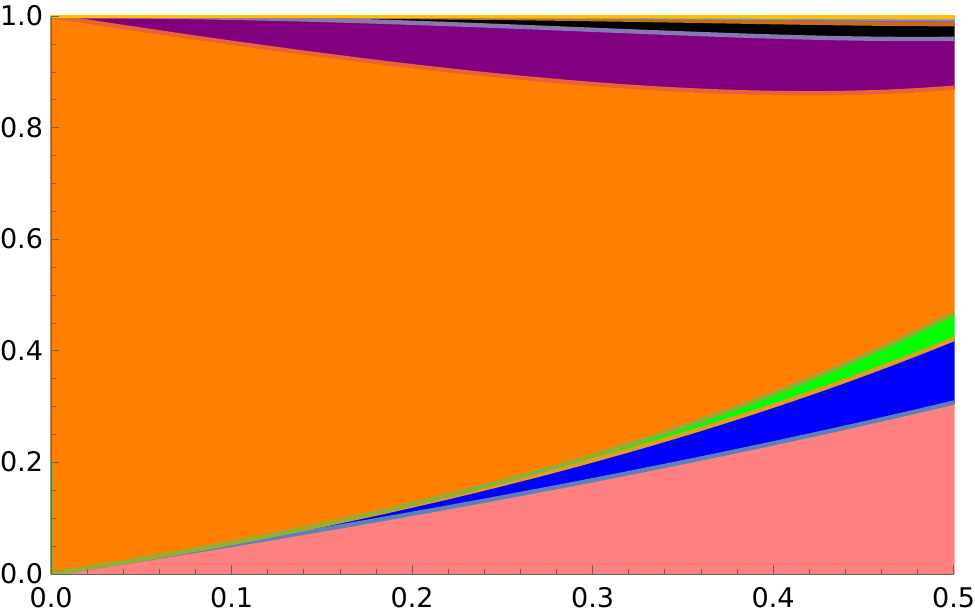}
    \caption{Initial type $j=3$.} 
 \end{subfigure}
  \caption{Average type proportions \eqref{fjkld3-2}
  as functions of $p\in [0,1/2)$. %
} 
\label{fig2-2} 
\end{figure}

\vspace{-0.3cm}

\begin{figure}[H]
  \centering
 \begin{subfigure}[b]{0.45\textwidth}
    \includegraphics[width=1\linewidth, height=3.9cm]{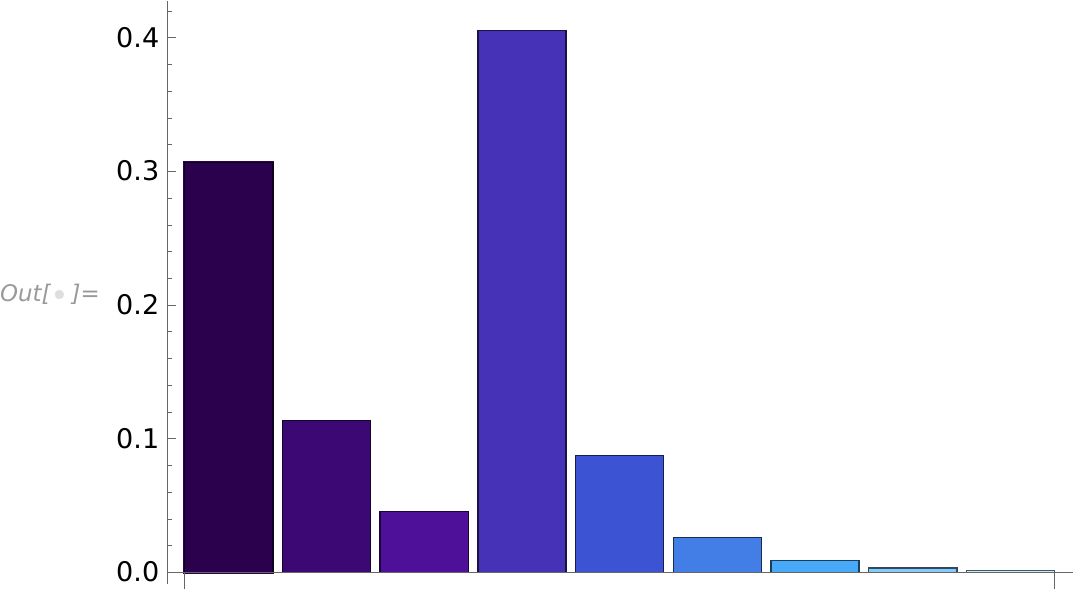}
    \caption{Initial type $j=3$.} 
 \end{subfigure}
 \begin{subfigure}[b]{0.45\textwidth}
    \includegraphics[width=1\linewidth, height=3.9cm]{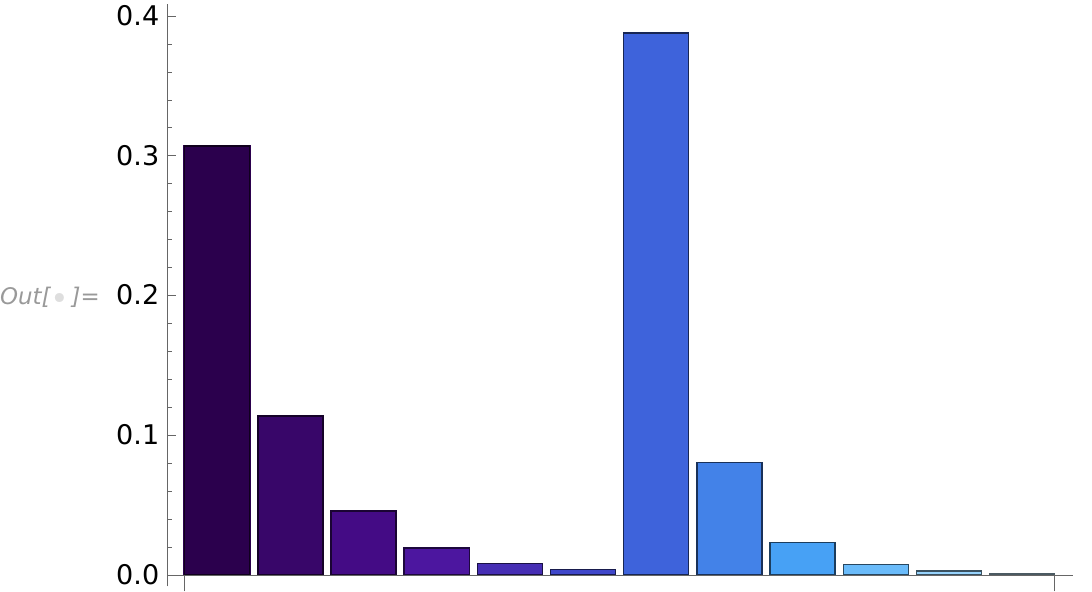}
    \caption{Initial type $j=6$.} 
 \end{subfigure}
\caption{Limiting distributions \eqref{jkld2} for $p=1/2$.
} 
\label{fig2-2-1} 
\end{figure}

\noindent 
 Corollary~\ref{fjklds13-1} 
 also yields the limiting values of
 the mean proportions \eqref{fjkld3-2} 
 as $p$ tends to $1/2$, i.e.
\begin{align}
  \nonumber
  & \lim_{p\to 1/2}
 \E_j \left[ \frac{X^{(k)}}{
        S^{\scaleto{\neq 0}{6pt}}_\infty }
      \ \! \Big| \ \!
      S^{\scaleto{\neq 0}{6pt}}_\infty \geq 1
      \right]
 \\
 \label{jkld2}
 & \quad \qquad
  =
 {\rm B} \left( \frac{1}{2},k+1,-1\right)
 +
 \ind_{\{k > j\}}
 \left( 
(k+1-j) B\left(
 \frac{1}{2},k-j,0
 \right)
 -2^{j-k+1}
 \right)
, 
\end{align} 
\noindent
as illustrated in Figure~\ref{fig2-2-1}.
  
\vspace{-0.3cm}

\subsection{Generating functions} %
\label{s3-1}
\noindent
{Let $F_0(p,r)=1$, $p,r\geq 0$, and for $n\geq 1$, }
    \begin{equation*}
    F_n(p,r)
    = \frac{r}{np+r} \binom{np+ r}{n}
    = \frac{r}{n} \binom{np+r-1}{n-1}
{= \frac{r \Gamma(np + r)}{\Gamma(n + 1) \, \Gamma((p-1)n + r + frac)}}
    ,
    \quad p,r \geq 0, 
    \end{equation*}
 denote the generalized
 Catalan numbers, or two-parameter Fuss--Catalan numbers,
 see \cite{Mlo10}. Then, the $n$-$th$ Catalan number is given
 by
 $$
 C_n = F_n ( 2,1) = \frac{1}{n+1}{2n\choose n},
 \quad n \geq 0.
$$ 
 In Proposition~\ref{B-special}
 we derive a closed-form conditional generating function
 expression using Fuss--Catalan numbers,
 which is proved in Appendix~\ref{s4}. 
\begin{prop}
\label{B-special}
  For any $\gamma \geq -1$ and $m \geq 1$, we have 
  \begin{equation}
\label{hhfdsjf} 
    \E_0 \left[
            \prod_{k=1}^{S^{\scaleto{\neq 0}{6pt}}_\infty}
        \left( 1 + \frac{\gamma}{k} \right)^{X^{(k)}}
   \ \! \Big| \ \!
        S^{\scaleto{\neq 0}{6pt}}_\infty = m
        \right]
    =
    \frac{F_m(\gamma + 2 , \gamma + 1)}{F_m(2 , 1)}. 
\end{equation}
\end{prop}
\noindent
         \noindent
 By differentiation of the generating
 function \eqref{hhfdsjf}, we have
 $$
      \E_0 \left[
            \sum_{k=1}^{S^{\scaleto{\neq 0}{6pt}}_\infty}
        \frac{X^{(k)}}{k} 
   \ \! \Big| \ \!
        S^{\scaleto{\neq 0}{6pt}}_\infty = m
        \right]
  = 
    \frac{\partial}{\partial \gamma}
   \frac{F_m(\gamma + 2 , \gamma + 1)}{F_m(2 , 1)}_{\big| \gamma = 0}
      =
    \sum_{j=1}^m 
    \frac{m+1}{m + j }, 
$$ 
hence
$$ 
 \lim_{m\to \infty} \frac{1}{m}   \E_0 \left[
            \sum_{k=1}^{S^{\scaleto{\neq 0}{6pt}}_\infty}
        \frac{X^{(k)}}{k} 
   \ \! \Big| \ \!
        S^{\scaleto{\neq 0}{6pt}}_\infty = m
        \right]
   = 
 \lim_{m\to \infty}
 \sum_{j=1}^m 
 \frac{1}{m + j }
  = \log 2. 
$$   
 The following corollary generalizes 
          \eqref{fjkldf34} from $\gamma = 0$ to any $\gamma \geq -1$. 
          \begin{corollary}
            \label{c1}
            The generating function
$$ 
  G^\gamma_0 ( \delta ) :=
\E_0 \left[
         \delta^{S^{\scaleto{\neq 0}{6pt}}_\infty} 
          \prod_{k=1}^{S^{\scaleto{\neq 0}{6pt}}_\infty} 
          \left( 1 + \frac{\gamma}{k} \right)^{X^{(k)}}
               \right]
$$
 solves the equation
 \begin{equation}
   \label{fjkld243} 
   \left( 1- \delta p G_0^\gamma(\delta ) \right)^{1+\gamma}
   G_0^\gamma(\delta ) = q. 
\end{equation}
          \end{corollary}
\begin{Proof} %
  From Propositions~\ref{radius-conv-2} and
 \ref{B-special}, we have 
\begin{align*}
  G^\gamma_0 ( \delta ) &:=
\E_0 \left[
         \delta^{S^{\scaleto{\neq 0}{6pt}}_\infty} 
          \prod_{k=1}^{S^{\scaleto{\neq 0}{6pt}}_\infty} 
          \left( 1 + \frac{\gamma}{k} \right)^{X^{(k)}}
               \right]
    \\
    &= \sum_{m=0}^\infty
    \delta^m \P \big( S^{\scaleto{\neq 0}{6pt}}_\infty = m \big) 
        \E_0 \left[
            \prod_{k=1}^{S^{\scaleto{\neq 0}{6pt}}_\infty}
        \left( 1 + \frac{\gamma}{k} \right)^{X^{(k)}}
   \ \! \Big| \ \!
        S^{\scaleto{\neq 0}{6pt}}_\infty = m
        \right]
        \\
        &= q \sum_{m=0}^\infty (pq \delta )^m 
  F_m(\gamma + 2 , \gamma + 1)
    \\
     & = \frac{1}{p\delta }\Phi_\gamma^{-1} ( pq \delta ) 
\end{align*}
 by Lemma~\ref{A2} below,
 where $\Phi^{-1}_\gamma$ the inverse function of 
\begin{equation}
\nonumber %
  \Phi_\gamma (w) := w(1-w)^{1+\gamma}, \quad w\in\mathbb C, 
 \end{equation}
 which yields \eqref{fjkld243}. 
\end{Proof}
                    \noindent 
          For example,
 taking $\gamma=1$, \eqref{fjkld243} 
 becomes a cubic equation that can be solved in closed form as
\begin{align*} 
\E_0 \left[
         \delta^{S^{\scaleto{\neq 0}{6pt}}_\infty} 
          \prod_{k=1}^{S^{\scaleto{\neq 0}{6pt}}_\infty} 
          \left( 1 + \frac{1}{k} \right)^{X^{(k)}}
               \right]
& = \frac{2}{3p \delta}
          - 
          \frac{1}{
            3\times 2^{2/3}\big( 27 \delta^4 p^4 q -2\delta^3 p^3 + 
     3 \sqrt{3\delta^7 p^7 q (27 \delta p q -4)}
     \big)^{1/3}}
          \\
           & \quad - \frac{\big(
   ( 27 \delta^4 p^4 q -2 \delta^3 p^3 + 
   3 \sqrt{3\delta^7 p^7 q (27 \delta p q-4)}\big)^{1/3}}{6\times 2^{1/3}\delta^2 p^2}. 
\end{align*}
\noindent
As a consequence of Proposition~\ref{B-special}, we also obtain
the following integrability criterion for product functionals.
\begin{corollary}
\label{c2} 
 Let $\delta >0$ and $\gamma \geq -1$,
 and let $(\sigma (k))_{k\geq 0}$ be a real sequence such that 
  \begin{equation}
      \label{fjkl34-0} 
      0 \leq \sigma (0) <
      \frac{(1+\gamma)^{1+\gamma}}{(2+\gamma)^{2+\gamma}pq\delta},
        \quad
  \mbox{and}
  \quad 
   0 \leq \sigma (k) \leq \left( 1+ \frac{\gamma}{k} \right)\delta , \quad k\geq 1. 
\end{equation} 
 Then, we have 
$$ 
 \E_{ {0}} \left[ 
   \sigma(0)^{S^{\scaleto{\neq 0}{6pt}}_\infty}
   \prod_{k=1}^{S^{\scaleto{\neq 0}{6pt}}_\infty}
    \sigma (k)^{X^{(k)}}
 \right] 
 < \infty. 
$$
\end{corollary}
\begin{Proof} %
{By Proposition~\ref{B-special}, we have} 
\begin{align}
 \nonumber
 &   \E_0 \left[
 ( \sigma(0) \delta )^{S^{\scaleto{\neq 0}{6pt}}_\infty}
      \prod_{k=1}^{S^{\scaleto{\neq 0}{6pt}}_\infty}
    \left( 1 + \frac{\gamma}{k} \right)^{X^{(k)}}
             \right] 
 = 
  \sum_{m=0}^\infty 
 { ( \sigma(0) \delta )^m}
\E_0 \left[
  { \prod_{k=1}^m \left( 1 + \frac{\gamma}{k} \right)^{X^{(k)}
    }
    }
   \ \! \Big| \ \!
   S^{\scaleto{\neq 0}{6pt}}_\infty = m
   \right]
    \P_0 \big( S^{\scaleto{\neq 0}{6pt}}_\infty = m \big) 
   \\
   \nonumber
   & \qquad
    = 
 q %
 \sum_{m=0}^\infty (pq )^m
 { ( \sigma(0) \delta )^m } 
 C_m
    \E_0 \left[
   \prod_{k=1}^{{m}} \left( 1 + \frac{\gamma}{k} \right)^{X^{(k)}
    }
   \ \! \Big| \ \! 
   S^{\scaleto{\neq 0}{6pt}}_\infty = m
   \right]
    \\
    \label{series}
  & \qquad = q %
  \sum_{m=0}^\infty (pq \sigma(0) {\delta } )^m 
  F_m(\gamma + 2 , \gamma + 1). 
\end{align}
 From the relation ${\Gamma(x+\alpha)} / {\Gamma(x)} = O(x^\alpha)$,
 we obtain   
\begin{align*}
  \limsup_{m\to\infty} \frac{
    F_{m+1}(\gamma + 2 , \gamma + 1)
  }{
    F_m(\gamma + 2 , \gamma + 1)
    } &= \limsup_{m\to\infty} \frac{\Gamma((2+\gamma)(m+1)+ \gamma+1) \Gamma((1+\gamma)(m+1))}{(m+2) \Gamma((1+\gamma)(m+2)) \Gamma((2+\gamma)m+ \gamma+1)}
  \\
    &= \limsup_{m\to\infty} \frac{
      ( (2+\gamma)m+ \gamma+1 )^{2+\gamma}}{(m+2)
    ( (1+\gamma)(m+1) )^{1+\gamma}}
  \\
   & = \frac{(2+\gamma)^{2+\gamma}}{(1+\gamma)^{1+\gamma}}, 
\end{align*}
 hence under \eqref{fjkl34-0} we have 
  \begin{equation}
\nonumber %
   \limsup_{m\to\infty} \frac{F_{m+1}(\gamma + 2 , \gamma + 1)}{
     F_m(\gamma + 2 , \gamma + 1)} <
         \frac{1}{pq\sigma(0) {\delta }}, 
\end{equation}
 and the series \eqref{series} converges absolutely. 
 {We conclude from the inequality} 
$$ 
    \E_0 \left[
 \sigma(0)^{S^{\scaleto{\neq 0}{6pt}}_\infty}
      \prod_{k=1}^{S^{\scaleto{\neq 0}{6pt}}_\infty}
    \sigma(k)^{X^{(k)}}
             \right] 
   \leq
       \E_0 \left[
 ( \sigma(0) \delta )^{S^{\scaleto{\neq 0}{6pt}}_\infty}
      \prod_{k=1}^{S^{\scaleto{\neq 0}{6pt}}_\infty}
    \left( 1 + \frac{\gamma}{k} \right)^{X^{(k)}}
             \right] 
       = G_0^\gamma(\sigma (0) \delta ),
$$ 
 {that follows from \eqref{fjkl34-0}.} 
\end{Proof}
\noindent

\section{Continuous-time setting}
\label{s3}
\subsection{Marked binary branching process} %
\noindent
In this section, we consider an age-dependent continuous-time
random tree %
${\cal T}_t$, $t>0$, in which
 the lifetimes of branches 
 are independent and identically distributed via a 
 common exponential density function
 $\rho (t) = \lambda e^{-\lambda t}$, $t\geq 0$, 
 with parameter $\lambda > 0$. 
 In addition to a type $j\in \inte$, %
 a label $\mathbf{k}$ in 
$$
 \mathbb K := \{ \varnothing  \}\cup \bigcup_{n\geq
  {1} } \{1,2\}^n,
$$ 
 is attached to every branch, %
 as follows. 
\begin{itemize}
\item At time $0$ we start from a
  single branch with label $\mathbf{k}=\varnothing$
  and initial type $j\in \inte$.   
 At the end of its lifetime $T_\varnothing$, 
 this branch yields either: 
 \begin{itemize}
\item
 no offspring if $T_\varnothing \geq t$; 
\item
 two independent offsprings
 with respective labels $(1)$, $(2)$
 and respective types $0$, $j+1$
 if $T_\varnothing < t$. %
\end{itemize}
 \item
  At generation $n \geq 1$, a branch 
  having a parent label 
  $\mathbf{k}- := \left(k_1, \ldots, k_{n-1}\right)$ 
  and type $i\in \inte$ starts at time $T_{{\mathbf k}-}$
  and has the lifetime $\tau_{\mathbf k}$.
 At the end of its lifetime
 $T_{\mathbf k} := T_{{\mathbf k}-} + \tau_{\mathbf k}$, 
 this branch yields either: 
\begin{itemize}
\item
 no offspring if $T_{\mathbf k} \geq t$; 
\item
 two independent offsprings
 with respective labels 
 $(\mathbf{k},1) = (k_1, \ldots , k_n,1)$
 and 
 $(\mathbf{k},2) = (k_1, \ldots , k_n,2)$, 
 and respective types $0$, $i+1$ if $T_{\mathbf k} < t$;
\end{itemize}
\end{itemize}
\noindent 
 see Figure~\ref{fig3}. %
 In particular, when a branch $\mathbf k$
 with type $i\geq 0$ splits,
 its two offsprings are respectively marked by $0$ and $i+1$. 
 
\begin{figure}[H]
\centering
\tikzstyle{level 1}=[level distance=6cm, sibling distance=4cm]
\tikzstyle{level 2}=[level distance=6cm, sibling distance=5cm]
\tikzstyle{level 2}=[level distance=6cm, sibling distance=4cm]
\begin{center}
\resizebox{0.98\textwidth}{!}{
\begin{tikzpicture}[scale=1.0,grow=right, sloped]
        \node[rectangle,draw,black,text=black, font=\Large,thick,fill=gray!15] {\large $0$}
            child {
  node[name=data, %
    draw, font=\Large, rounded corners=5pt, fill=gray!15,xshift=-0.5cm,
  ] (main) 
        {\large $T_\varnothing$} %
                child{
                  node[%
                    draw,black,text=black, font=\Large,thick,yshift=-1.4cm,xshift=7cm,rounded corners=5pt, fill=gray!15]{\large $T_{(2)}$} %
                    child{
    node[name=data, %
         draw,yshift=-0.6cm, font=\Large, rounded corners=5pt,yshift=0cm,xshift=7cm, fill=gray!15]  
        {$t$} %
                    edge from parent
                    node[above,trapezium, font=\Large,draw,fill=gray!15]{$(2,2)$}
                    node[rectangle,draw,black,text=black, font=\Large, font=\Large, thick,fill=purple!60,below]{$j+2=5$}
                    }
                    child{
    node[name=data, %
         draw, font=\Large, rounded corners=5pt,yshift=-0.6cm,xshift=7cm,fill=gray!15]  
        {$t$} %
                    edge from parent
                    node[above,trapezium, font=\Large,draw,fill=gray!15]{$(2,1)$}
                    node[rectangle,draw,black,text=black, font=\Large,thick,fill=gray!15,below]{ $0$}
                    }
                edge from parent
                node[above,trapezium, font=\Large,draw,fill=gray!15]{$(2)$}
                node[rectangle,draw,black,text=black, font=\Large,thick,fill=orange!60,below]{ $j+1=4$}
                }
                child{
                  node[%
                    draw,black,text=black, font=\Large,thick,yshift=0.4cm,rounded corners=5pt,fill=gray!15]{\large $T_{(1)}$} %
                    child{
                      node[%
                        draw,black,text=black, font=\Large,thick,yshift=0.4cm,xshift=-1cm,rounded corners=5pt,fill=gray!15]{\large $T_{(1,2)}$} %
                    child{
    node[name=data,  %
         draw, font=\Large, rounded corners=5pt,yshift=0.6cm,xshift=9cm,fill=gray!15]  
        {$t$} %
                    edge from parent
                    node[above,trapezium, font=\Large,draw,fill=gray!15]{$(1,2,2)$}
                    node[rectangle,draw,black,text=black, font=\Large,thick,fill=blue!40,below]{ $2$}
                    }
                    child{
                      node[%
                        draw,black,text=black, font=\Large,thick,yshift=0.4cm,xshift=3cm,rounded corners=5pt,fill=gray!15]{\large $T_{(1,2)}$} %
                    child{
    node[name=data, %
         draw, font=\Large, rounded corners=5pt,yshift=0cm,fill=gray!15]  
        {$t$} %
                    edge from parent
                    node[above,trapezium, font=\Large,draw,fill=gray!15]{$(1,2,1,2)$}
                    node[rectangle,draw,black,text=black, font=\Large,thick,fill=pink,below]{ $1$}
                    }
                    child{
    node[name=data, %
         draw, font=\Large, rounded corners=5pt,yshift=-0.6cm,fill=gray!15]  
        {$t$} %
                    edge from parent
                    node[above,trapezium, font=\Large,draw,fill=gray!15]{$(1,2,1,1)$}
                    node[rectangle,draw,black,text=black, font=\Large,thick,fill=gray!15,below]{ $0$}
                    }
                edge from parent
                node[above,trapezium, font=\Large,draw,fill=gray!15]{$(1,2,1)$}
                node[rectangle,draw,black,text=black, font=\Large,thick,fill=gray!15,below]{ $0$}
                    }
                                        edge from parent
                    node[above,trapezium, font=\Large,draw,fill=gray!15]{$(1,2)$}
                    node[rectangle,draw,black,text=black, font=\Large,thick,fill=pink,below]{ $1$}
}
                    child{
    node[name=data, font=\Large, %
      draw, yshift=2cm, xshift=14cm,
      rounded corners=5pt,fill=gray!15]  
        {$t$} %
                    edge from parent
                    node[above,trapezium, font=\Large,draw,fill=gray!15]{$(1,1)$}
                    node[rectangle,draw,black,text=black, font=\Large,thick,fill=gray!15,below]{$0$}
                    }
                edge from parent
                node[above,trapezium, font=\Large,draw,fill=gray!15]{$(1)$}
                node[rectangle,draw,black,text=black, font=\Large,thick,fill=gray!15,below]{ $0$}
                }
                edge from parent
                node[above,trapezium, font=\Large,draw,fill=gray!15] {$\varnothing$}
                node[rectangle,draw,black,text=black, font=\Large,thick,fill=green!60,below]{ $j=3$}
            };
\end{tikzpicture}
}
\end{center}
\vskip-0.2cm
\caption{Sample of the marked random tree ${\cal T}_t$,
  $t >0$, started from the initial type $j=3$.}
\label{fig3} 
\end{figure}

\vskip-0.3cm

\noindent
\noindent 
 We refer to e.g. \cite[Eq. (8) page 3]{kendall1948},
 \cite[Example 13.2 page 112]{harris1963}, 
 and \cite[Example 5 page 109]{athreya} 
 or \cite[Equation~(5)]{iwasa} 
 for the following result, whose proof is given
 in Appendix~\ref{s5} for completeness. 
\begin{prop} 
\label{progeny}
 The distribution of the count $N_t$ 
 of nodes with non-zero types in ${\cal T}_t$, $t\geq 0$,
 {excluding the initial node,} %
 is given by 
 \begin{equation}
 \label{recursion-total-sol}
  {\P}( N_t = m )
  =
  e^{-\lambda t} (1-e^{-\lambda t})^m, \quad m\geq 0,
\end{equation}
 and probability generating function 
\begin{equation}
\label{pgf-ct}
G_t(\delta ) = \E \big[ \delta^{N_t} \big] 
= \frac{%
  {e^{-\lambda t}}
}{1 - (1-e^{-\lambda t}) {\delta }}, \quad
 t\geq 0. 
\end{equation}
\end{prop}
\noindent 
 In addition, it follows from \eqref{pgf-ct} that
 $\E[N_t ] = e^{\lambda t}-1$, $t\geq 0$. 
 Note also that $N_t$ equals the number of nodes with type $0$
 excluding the initial node, since each node with non-zero type
 has a co-twin with type $0$.
 Hence, the total progeny of the random tree
 ${\cal T}_t$, $t\geq 0$, is $2N_t+1$.
\subsection{Conditional type distribution} 
\noindent
 In what follows, we let $X_t^{(i)}$ denote the count of types
 equal to $i \geq 1$ until time $t$,
 {excluding the initial node}. %
 
 \medskip

\begin{figure}[H]
  \centering
 \begin{subfigure}[b]{0.45\textwidth}
    \includegraphics[width=1\linewidth, height=3.9cm]{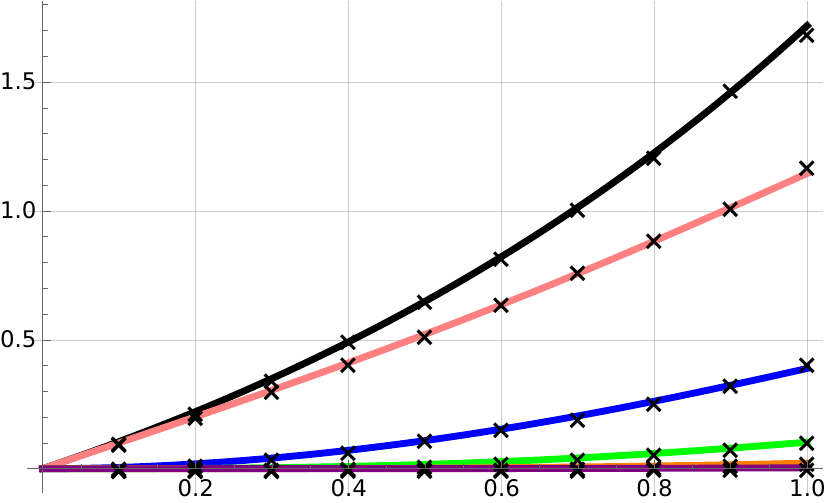}
    \caption{Initial type $j=0$.} 
 \end{subfigure}
 \begin{subfigure}[b]{0.45\textwidth}
    \includegraphics[width=1\linewidth, height=3.9cm]{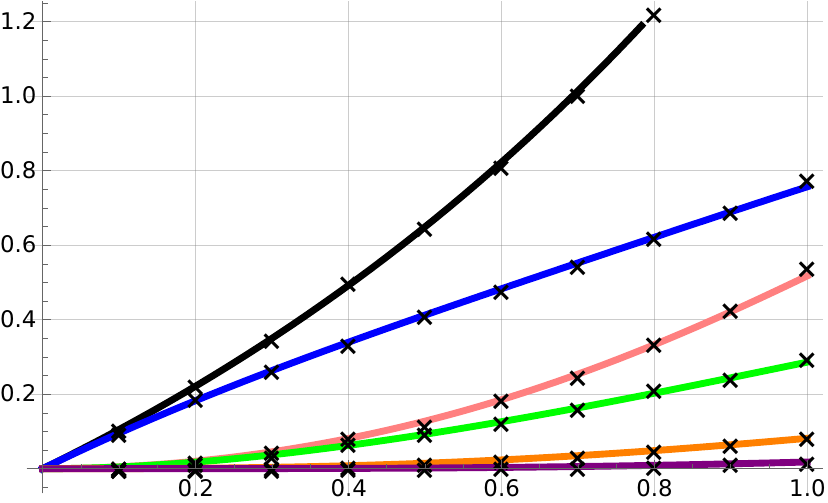}
    \caption{Initial type $j=1$.} 
 \end{subfigure}
 \begin{subfigure}[b]{0.45\textwidth}
    \includegraphics[width=1.0\linewidth, height=3.9cm]{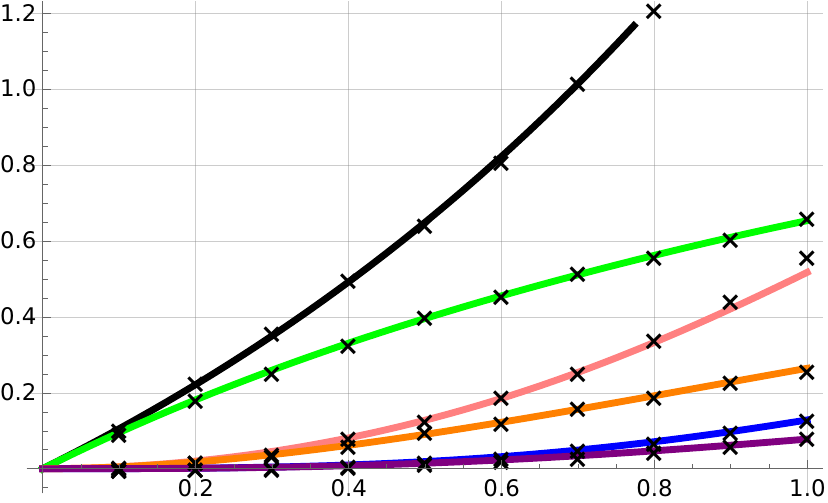}
    \caption{Initial type $j=2$.} 
 \end{subfigure}
  \centering
  \begin{subfigure}[b]{0.45\textwidth}
 \hskip0.1cm
    \includegraphics[width=1.0\linewidth, height=3.9cm]{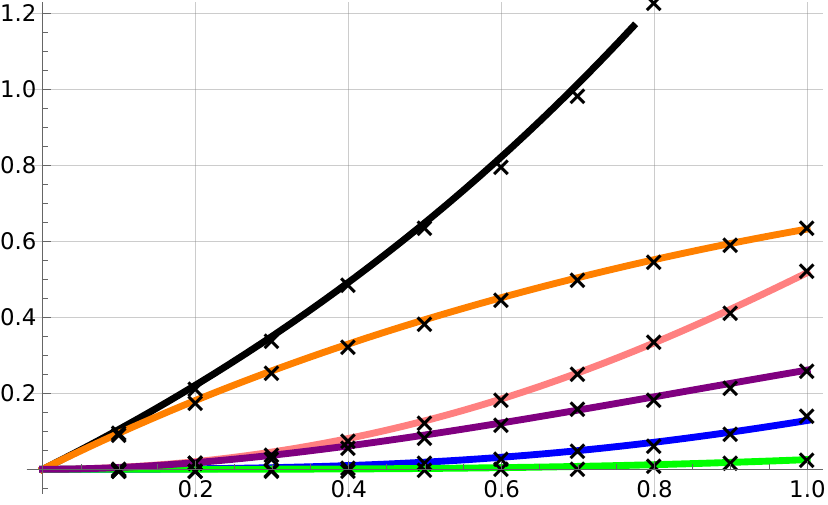}
    \caption{Initial type $j=3$.} 
 \end{subfigure}
  \caption{Expected counts \eqref{fjkldf1-0} of types
      as functions of $t\in [0,1]$
  with $\lambda =1$.} 
\label{fig2-11-2-0} 
\end{figure}

\begin{figure}[H]
  \centering
 \begin{subfigure}[b]{0.45\textwidth}
    \includegraphics[width=1\linewidth, height=3.9cm]{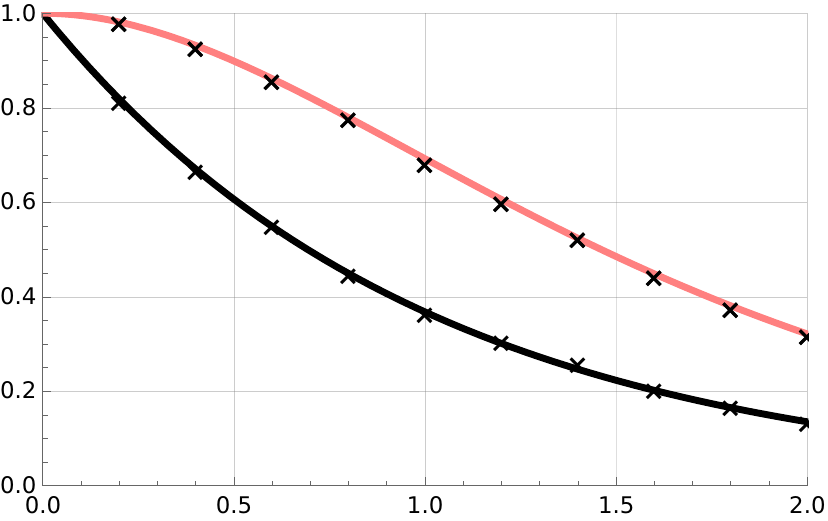}
    \caption{Initial type $j=0$.} 
 \end{subfigure}
 \begin{subfigure}[b]{0.45\textwidth}
    \includegraphics[width=1\linewidth, height=3.9cm]{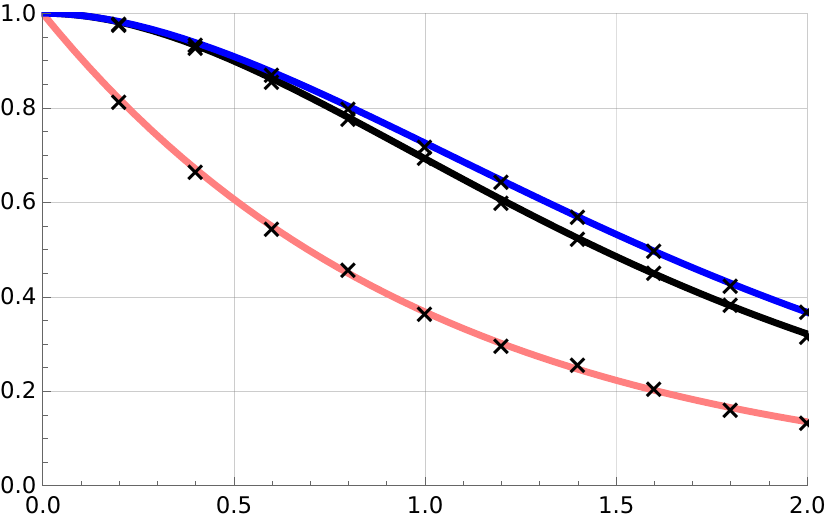}
    \caption{Initial type $j=1$.} 
 \end{subfigure}
 \begin{subfigure}[b]{0.45\textwidth}
    \includegraphics[width=1.0\linewidth, height=3.9cm]{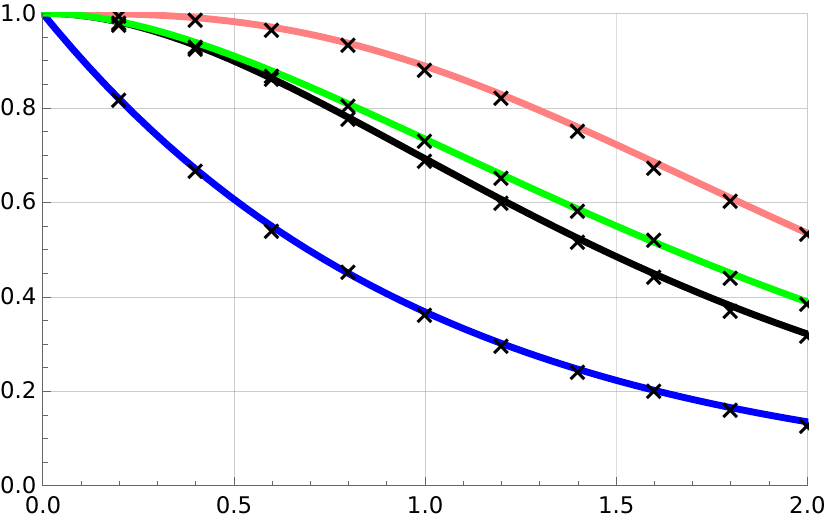}
    \caption{Initial type $j=2$.} 
 \end{subfigure}
  \centering
  \begin{subfigure}[b]{0.45\textwidth}
 \hskip0.1cm
    \includegraphics[width=1.0\linewidth, height=3.9cm]{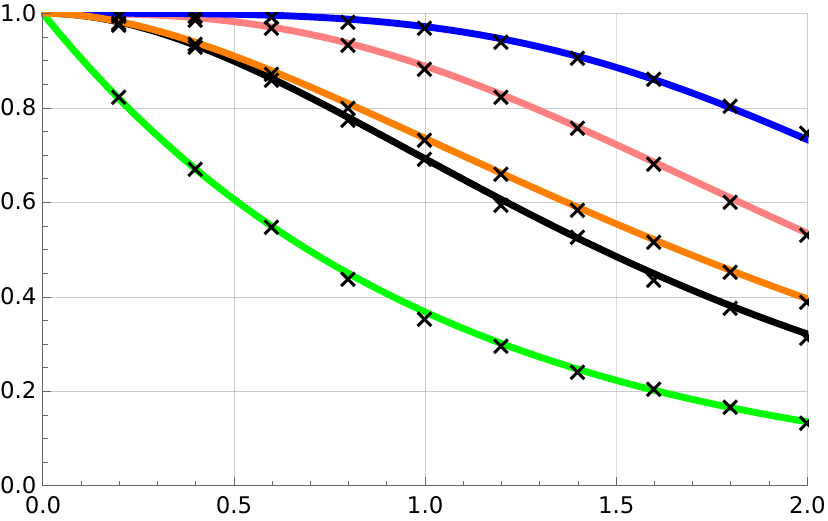}
    \caption{Initial type $j=3$.} 
 \end{subfigure}
  \caption{Tail CDFs \eqref{fjkldf1-04} of the occurrence times
    of given types
   with $\lambda =1$.} 
\label{fig2-11-2-0-2} 
\end{figure}

 In Theorem~\ref{const-l-weight}, 
 which is proved in Appendix~\ref{s5},
 we compute recursively
 the conditional type distribution
 of $\big( X^{(1)}_t ,\ldots , X^{(n)}_t\big)$ given
 that their summation
 equals $N_t$ and $X^{(k)}_t =0$ for all $k>n$,
 and show that it does
 not depend on time $t>0$ and on the parameter $\lambda >0$. 
\begin{theorem}
\label{const-l-weight}
For $j \geq 0$
{and $n\geq 1$}, the conditional type distribution 
\begin{equation}
\label{recursion-1-sol-j}
  a_j (m_1,\ldots,m_n)
  :=
  \P_j \big( X_t^{(1)} = m_1,\ldots , X_t^{(n)}=m_n
  \mid N_t = m_1+\cdots + m_n \big) 
\end{equation}
  is given by the recursion 
\begin{equation}
  \label{recursion-1-coef-general-j}
  a_j (m_1,\ldots,m_n) = \sum_{l=1}^{n-j}
  \frac{1}{l!}
  \ind_{\{
      m_{j+l} > m_{j+l+1}
      \}}
  \hskip-0.5cm
  \sum_{\begin{subarray}{c}
      m_i^1+\cdots + m_i^l = m_i -\ind_{\{j < i\leq j+l\}}, \ \! 1\leq i\leq n \\
      0\leq m^k_i \leq m_{i-1}^k, \ 2\leq i\leq n, \ \! 1\leq k\leq l
  \end{subarray}} \prod_{k=1}^l \frac{a_0( m^k_1,\ldots,m^k_n )}{1 +
    m^k_1+\cdots + m^k_n },
\end{equation}
$m_1,\ldots , m_n\geq 0$, with
{$m_{n+1} := 0$ in
  the last indicator
  $\ind_{\{
      m_n > m_{n+1}
      \}}$,}
$a_j (\varnothing) := 1$, 
$a_j (m_1,\ldots ,m_n) = a_j (m_1,\ldots ,m_{n-1})$ if $m_n = 0$, 
and
$a_j (m_1,\ldots ,m_n) = 0$ if $1 \leq n < j$,
{$m_1 + \cdots + m_n \geq 1$}. 
\end{theorem} 
\noindent 
{From Theorem~\ref{const-l-weight},
 the conditional type distribution \eqref{recursion-1-sol-j} 
 can be computed by first applying the recursion 
 \eqref{recursion-1-coef-general-j} to $j=0$, as
$$
  a_0 (m_1,\ldots,m_n) = \sum_{l=1}^n
  \frac{1}{l!}
  \ind_{\{
      m_l > m_{l+1}
      \}}
  \hskip-0.5cm
  \sum_{\begin{subarray}{c}
      m_i^1+\cdots + m_i^l = m_i -\ind_{\{1\leq i\leq l\}}, \ \! 1\leq i\leq n \\
      0\leq m^k_i \leq m_{i-1}^k, \ 2\leq i\leq n, \ \! 1\leq k\leq l
  \end{subarray}} \prod_{k=1}^l \frac{a_0( m^k_1,\ldots,m^k_n )}{1 +
    m^k_1+\cdots + m^k_n },
$$ 
 and then applying again 
 \eqref{recursion-1-coef-general-j} 
 to $j\geq 1$. 
 Note also that when
 $m_1+\cdots + m_n < n - j$,
 the summation range in
 \eqref{recursion-1-coef-general-j} 
 is empty, whence
 $a_j (m_1,\ldots , m_n) = 0$.
 In addition,} for $j\geq 0$, $m\geq 1$,
  $1 \leq n \leq m+j$ 
  and any weight function $f_n : \inte^n \to \real$,
  we have 
\begin{equation}
  \label{fjklds13}
  \E_j \big[
   f_n \big( X^{(1)}_t,\ldots , X^{(n)}_t \big)
   {\bf 1}_{\{
     X^{(1)}_t + \cdots + X^{(n)}_t = m %
     \}}      
   \ \! \big| \ \!
   N_t = m
   \big]
 =
 \hskip-0.5cm
 \sum_{\begin{subarray}{c}
    (m_1,\ldots,m_n)\in {\mathbb K}_{j,n} \\
    m_1 + \cdots + m_n = m
    \end{subarray}}
 \hskip-0.3cm
 a_j(m_1,\ldots,m_n)
 f_n (m_1,\ldots , m_n). 
\end{equation}
\noindent 
 Figure~\ref{fig2-11-2-0} displays the time evolution of the expected values 
\begin{equation}
  \label{fjkldf1-0}
   \E_j \big[ X_t^{(l)} \big] =
 \sum_{m=1}^\infty
 \E_j \big[ X_t^{(l)} \ \! \big| \ \! N_t = m \big]
 \P ( N_t = m )
\end{equation} 
of the count of non-zero types computed as functions of $t\in [0,1]$ from
\begin{equation}
\label{fjkld3-22}
  \E_j \big[ X^{(l)}_t \ \! \big| \ \! N_t = m \big] 
 =
 \sum_{n=\max ( l  ,j)}^{m+j}
  \sum_{\begin{subarray}{c}
    (m_1,\ldots,m_n)\in {\mathbb K}_{j,n} \\
    m_1 + \cdots + m_n = m
    \end{subarray}}
  m_l a_j(m_1,\ldots,m_n),
  \quad l =1,\ldots , m + j, 
\end{equation}
\noindent
 by truncation of the series \eqref{fjkldf1-0} 
 up to $m=12$, after starting from the initial types
 $j=0,1,2,3$, 
 together with Monte Carlo simulations over 10,000 samples.
 Color codings are consistent with \eqref{dkjl} and those of
 Figures~\ref{fig1}-\ref{fig2-2} and \ref{fig3}. 
\noindent 
 Figure~\ref{fig2-11-2-0-2} displays the tail cumulative distribution functions 
\begin{equation}
  \label{fjkldf1-04}
  \P_j \big( \tau^{(l)} >t \big) =
   \P_j \big( X_t^{(l)} = 0 \big) =
   \P ( N_t = 0 )
   + \sum_{m=1}^\infty
 \P_j \big( X_t^{(l)} = 0 \ \! \big| \ \! N_t = m \big)
 \P ( N_t = m )
\end{equation} 
of the first time $\tau^{{(l)}}$ of occurrence of type $l$ 
 which, according to \eqref{recursion-1-sol-j}, is computed as 
\begin{align}
\label{fjkld3-224}
 &  \P_j \big( X^{(l)}_t = 0 \ \! \big| \ \! N_t = m \big) 
 = 
 \sum_{n=j}^{l-1}
  \sum_{\begin{subarray}{c}
    (m_1,\ldots,m_n)\in {\mathbb K}_{j,n} \\
    m_1 + \cdots + m_n = m
    \end{subarray}}
  a_j(m_1,\ldots,m_n)
  \\
  \nonumber 
   & 
 \ \ +
   \sum_{n=\max ( l  ,j)}^{m+j}
  \sum_{\begin{subarray}{c}
    (m_1,\ldots,m_{l-1},0,m_{l+1},\ldots , m_n)\in {\mathbb K}_{j,n} \\
    m_1 + \cdots + m_{l-1} + m_{l+1} + \cdots + m_n = m
    \end{subarray}}
  a_j(m_1,\ldots,m_{l-1},0,m_{l+1},\ldots , m_n),
  \quad l =1,\ldots , m + j. 
\end{align} 
\noindent
 For this, we truncate of the series \eqref{fjkldf1-04} 
 up to $m=12$, after starting from the initial types
 $j=0,1,2,3$. 
 The closed-form expressions are confirmed by
 Monte Carlo simulations over 10,000 samples,
 {and remain stable for $t\in [0,2$]}.

\begin{figure}[H]
  \centering
 \begin{subfigure}[b]{0.45\textwidth}
    \includegraphics[width=1\linewidth, height=3.9cm]{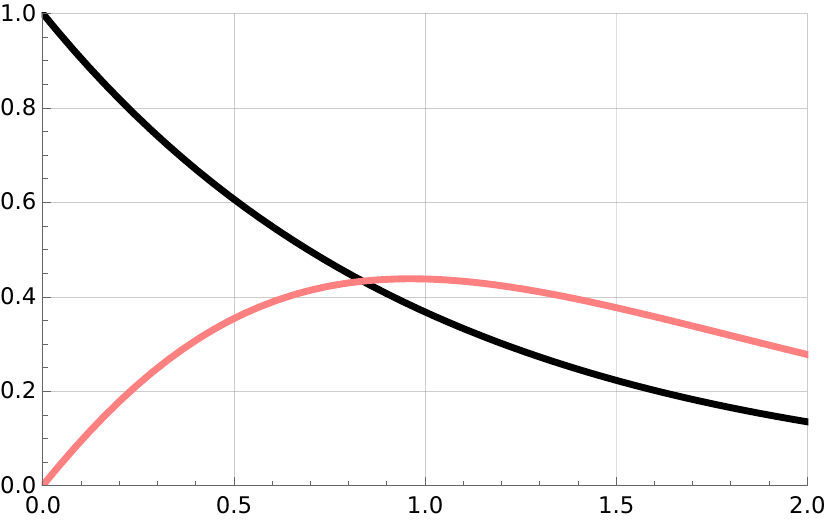}
    \caption{Initial type $j=0$.} 
 \end{subfigure}
 \begin{subfigure}[b]{0.45\textwidth}
    \includegraphics[width=1\linewidth, height=3.9cm]{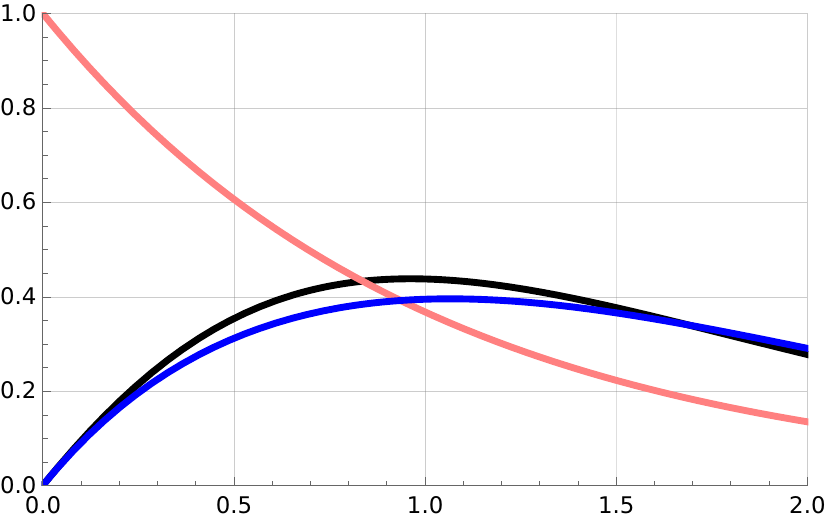}
    \caption{Initial type $j=1$.} 
 \end{subfigure}
 \begin{subfigure}[b]{0.45\textwidth}
    \includegraphics[width=1.0\linewidth, height=3.9cm]{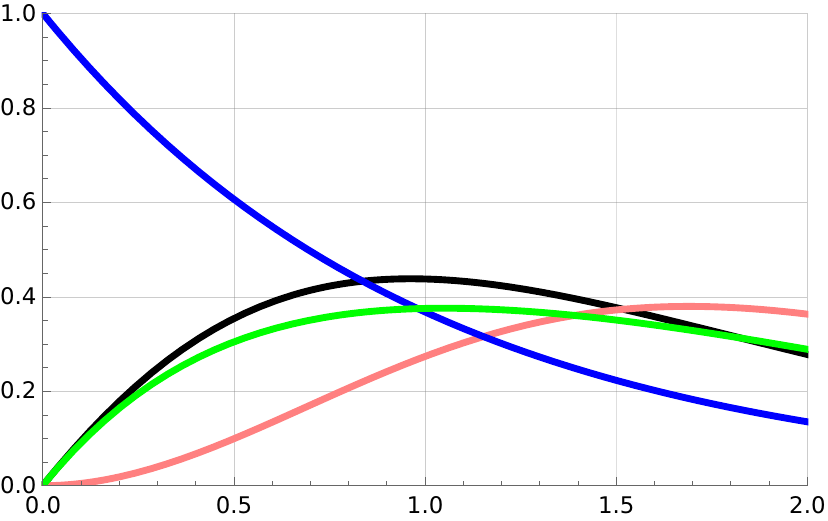}
    \caption{Initial type $j=2$.} 
 \end{subfigure}
  \centering
  \begin{subfigure}[b]{0.45\textwidth}
 \hskip0.1cm
    \includegraphics[width=1.0\linewidth, height=3.9cm]{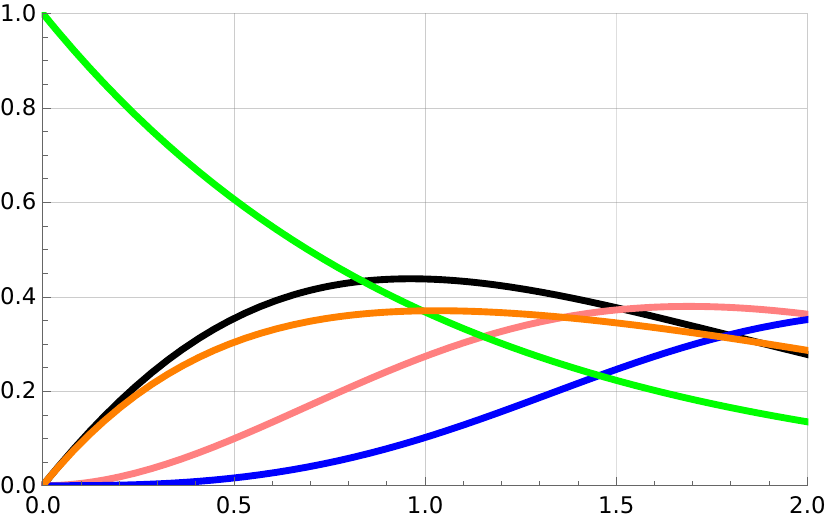}
    \caption{Initial type $j=3$.} 
 \end{subfigure}
  \caption{PDFs \eqref{fjkldf1-04} of the occurrence times
    of given types
   with $\lambda =1$.} 
\label{fig2-11-2-0-3} 
\end{figure}

\vskip-0.2cm 
\noindent
  Figure~\ref{fig2-11-2-0-3} displays the
  evolution of the probability density functions 
  of the first time $\tau^{{(l)}}$ of occurrence of type $l$,
  as obtained from
  \eqref{fjkldf1-04}
  and \eqref{fjkld3-224}
  for $t\in [0,2$], after starting from the initial types
  $j=0,1,2,3$. 

\begin{figure}[H]
  \centering
 \begin{subfigure}[b]{0.45\textwidth}
    \includegraphics[width=1\linewidth, height=3.9cm]{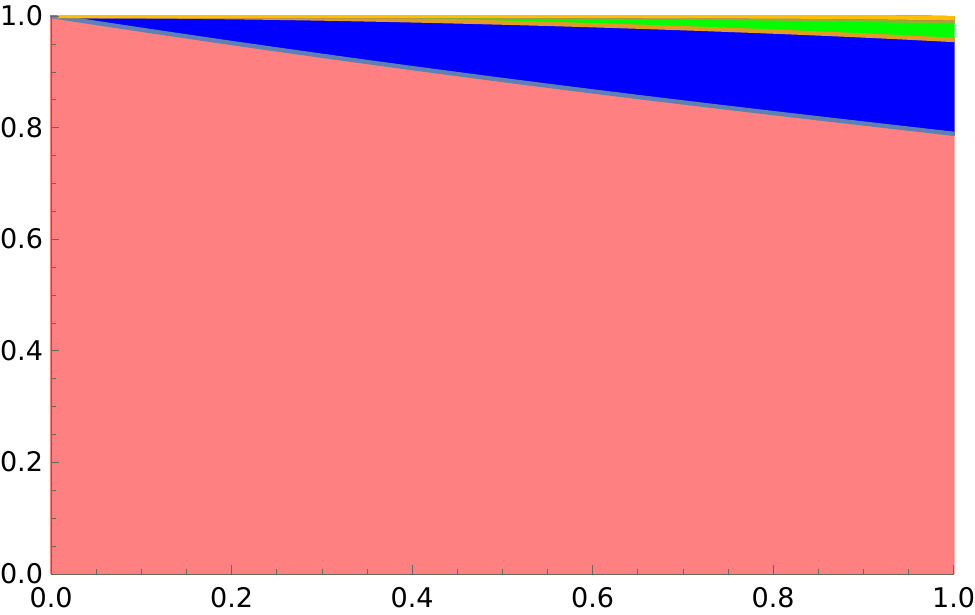}
    \caption{Initial type $j=0$.} 
 \end{subfigure}
 \begin{subfigure}[b]{0.45\textwidth}
    \includegraphics[width=1\linewidth, height=3.9cm]{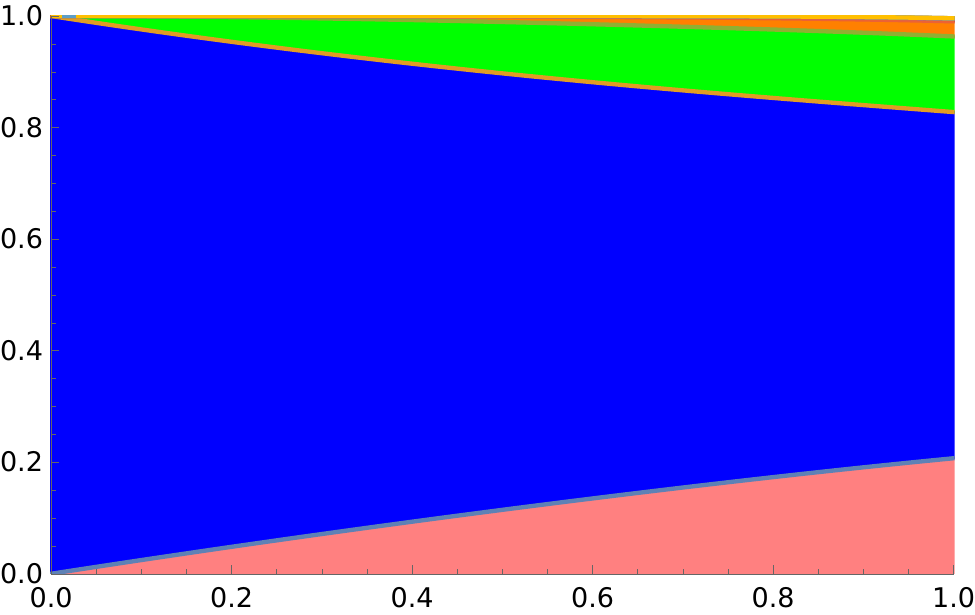}
    \caption{Initial type $j=1$.} 
 \end{subfigure}
 \begin{subfigure}[b]{0.45\textwidth}
    \includegraphics[width=1.0\linewidth, height=3.9cm]{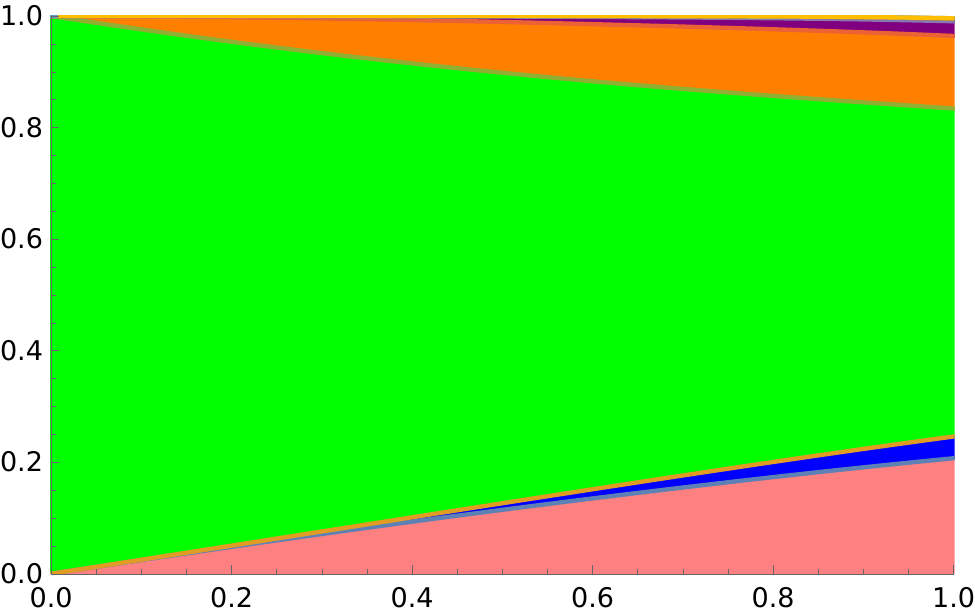}
    \caption{Initial type $j=2$.} 
 \end{subfigure}
  \centering
  \begin{subfigure}[b]{0.45\textwidth}
 \hskip0.1cm
    \includegraphics[width=1.0\linewidth, height=3.9cm]{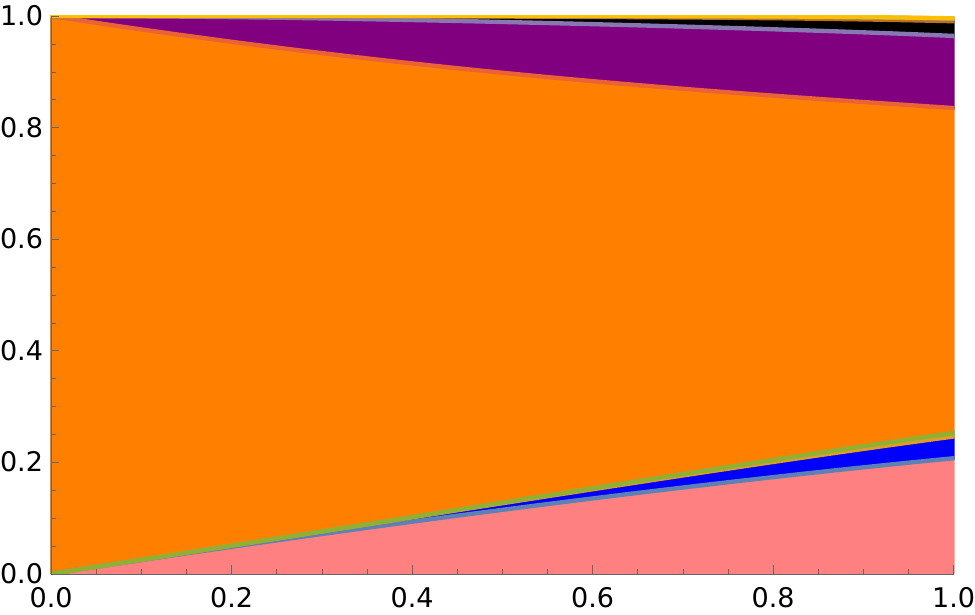}
    \caption{Initial type $j=3$.} 
 \end{subfigure}
  \caption{Mean proportions of types
    \eqref{fjkldf1}
    as functions of $t\in [0,1]$
  with $\lambda =1$.} 
\label{fig2-11-2} 
\end{figure}

 \noindent 
 Figure~\ref{fig2-11-2} displays the mean proportions
\begin{equation}
  \label{fjkldf1}
   \E_j \left[ \frac{X_t^{(l)}}{N_t} \ \! \Big| \ \! N_t \geq 1\right] =
 \frac{1}{1 - e^{-\lambda t}}
 \sum_{m=1}^\infty
 \frac{1}{m}
 \E_j \big[ X_t^{(l)} \ \! \big| \ \! N_t = m \big]
 \P ( N_t = m )
\end{equation} 
of non-zero types computed as
functions of $t\in (0,1)$ from
 \eqref{fjkld3-22}
 and truncation of the series \eqref{fjkldf1} 
 up to $m=12$, after starting from the initial types
 $j=0,1,2,3$. %
 Due to truncation, the computed proportions 
 are accurate and add up to 100\% only up to $t=1$. 
\subsection{Generating functions}
\label{s3-3}
\noindent 
In Proposition~\ref{hjkfd21}, which is proved in Appendix~\ref{s5}, 
 we derive a closed-form
 conditional generating function
 expression. 
\begin{prop}
  \label{hjkfd21}
  For any $\gamma , t>0$ and $m,j\geq 0$ we have 
   \begin{equation}
   \label{fjkl234} 
    \E_j \left[
      \prod_{k=1}^{N_t}
    (\gamma + k -2)^{X_t^{(k)}}
        \ \! \Big| \ \!
        N_t = m \right] 
   = (-\gamma )^m \binom{
      - 1 
      - ({j-1})/{\gamma}
   }{m}. 
\end{equation}
 \end{prop}
 \noindent
 In particular, for
 $j=0$, %
 $\gamma = 2$ and $t>0$, we have 
$$
  \E_0 \left[
      \prod_{k=1}^{N_t}
    k^{X_t^{(k)}}
        \ \! \Big| \ \!
        N_t = m \right] 
  = \frac{(2m)!}{2^m (m!)^2}, \quad m \geq 0. 
$$ 
 \begin{prop} 
  \label{A-special}
 For any $\delta , \gamma , t>0$ such that 
 $( 1 - e^{-\lambda t} ) \gamma \delta <1$, we have 
 \begin{equation}
   \label{fjkldf132} 
         \E_j \left[
           \delta^{N_t} 
           \prod_{k=1}^{N_t}
    (\gamma + k -2)^{X_t^{(k)}}
    \right] 
=
    \frac{e^{-\lambda t}}{
      ( 1- ( 1 - e^{-\lambda t} ) \gamma \delta )^{
      1 + (j-1) / \gamma 
    }}, 
    \quad j \geq 0. 
  \end{equation}
\end{prop} 
\begin{Proof} %
  By Propositions~\ref{progeny}
  and \ref{hjkfd21}, we have
   \begin{align*} 
         \E_j \left[
           \delta^{N_t} 
           \prod_{k=1}^{N_t}
    (\gamma + k -2)^{X_t^{(k)}}
    \right] 
 & = 
\sum_{m=0}^\infty
\P ( N_t = m )
\delta^m
    \E_j \left[
      \prod_{k=1}^m 
    (\gamma + k -2)^{X_t^{(k)}}
        \ \! \Big| \ \!
        N_t = m \right] 
    \\
 & = 
    e^{-\lambda t}
    \sum_{m=0}^\infty 
 ( 1 - e^{-\lambda t})^m (-\gamma \delta )^m \binom{
      - 1 
      - (j-1)/\gamma 
    }{m} 
    \\
&    =
    \frac{e^{-\lambda t}}{
      ( 1- ( 1 - e^{-\lambda t} ) \gamma \delta )^{
      1 + (j-1) / \gamma 
    }}
,
  \quad
  j\geq 0.
\end{align*} 
\end{Proof}
 \noindent
 In particular, for 
 $\delta = 1$, $\gamma = 2$ and $t>0$ we have 
  \begin{equation*}
         \E_j \left[
                    \prod_{k=1}^{N_t}
    k^{X_t^{(k)}}
    \right] 
=
    \frac{e^{-\lambda t}}{
      ( 2 e^{-\lambda t} -1)^{
      (j+1)/2
    }}, 
    \quad j \geq 0. 
  \end{equation*}
\noindent
As a consequence of Proposition~\ref{A-special}, we obtain
the following integrability criterion for product functionals.
\begin{corollary}
\label{c3} 
 Let $t> 0$, $j \geq 0$, $\delta >0$, $\gamma > 1$,
 and let $(\sigma (k))_{k\geq 0}$ be a real sequence such that 
    \begin{equation}
      \label{fjkl34} 
  0 \leq   \sigma (0) < \frac{1}{
      (1-e^{-\lambda t})\gamma \delta }
  \quad
  \mbox{and}
  \quad 
  0 \leq   \sigma (k) \leq (\gamma + k-2 )\delta , \quad k\geq 1. 
\end{equation} 
 Then, we have the bound 
$$ 
 \E_j \left[ 
   \sigma(0)^{N_t}
   \prod_{k=1}^{N_t}
    \sigma (k)^{X_t^{(k)}}
 \right] 
 \leq
 \frac{{e^{-\lambda t}} %
 }
  {
    ( 1- (1-e^{-\lambda t}) \gamma \delta \sigma (0) )^{
      1 + (j-1)/\gamma 
    }
    } < \infty.
$$
\end{corollary}
\begin{Proof} %
 By \eqref{fjkl34} we have 
$$ 
    \E_j \left[
 \sigma(0)^{N_t}
      \prod_{k=1}^{N_t}
    \sigma(k)^{X_t^{(k)}}
             \right] 
   \leq
       \E_j \left[
 ( \sigma(0) \delta )^{N_t}
      \prod_{k=1}^{N_t}
    (\gamma + k -2)^{X_t^{(k)}}
             \right] 
 ,
  \quad
  j\geq 0,
$$ 
 and we conclude from \eqref{fjkldf132}. %
\end{Proof}
\subsubsection*{Conclusion}
\noindent
 We have presented a multitype
 Galton--Watson {process} 
 that can model mutation and reversion in discrete and continuous time. 
 Through a recursive computation of 
 the joint distribution of types conditionally to the value of
 the total progeny, we have determined  
 the evolution of various expected quantities, such as the
 mean proportions of different types
 as the tree size or time increases,
 and the distribution of the first time of occurrence of
 a given type.
 In comparison with the literature on related multitype models,
 our approach does not rely on approximations. 
 
\appendix

\section{Proofs - discrete-time setting} 
\label{s4}
\begin{Proofy} {\em of Proposition~\ref{radius-conv-2}}.   
\noindent 
By \cite[Theorem 2]{Ott49},
the probability generating function $G$
of the total progeny $1+2 S^{\scaleto{\neq 0}{6pt}}_\infty$ of
{${\cal T}$} %
satisfies the quadratic equation
\begin{equation}
\nonumber %
G(\delta ) = \delta  q + \delta  p G(\delta )^2 
\end{equation}
 in a neighborhood of $0$, 
 and admits the solution \eqref{fjkldf34},
 in which the choice of minus sign follows from the initial condition
 $p_0 = \lim_{\delta \to 0} G(\delta ) = 0$.
 Letting $g(w) := q + p w^2$,
 by \cite[Corollary 3]{Ott49} we have
$\P\big( S^{\scaleto{\neq 0}{6pt}}_\infty<\infty\big) = 1$ if and only if $g'(1)\le1$,
i.e. $p \leq 1/2$, and 
\begin{align*} 
 \P\big( S^{\scaleto{\neq 0}{6pt}}_\infty < \infty\big) & =
 G(1)
 \\
 & = \frac{1-\sqrt{1-4 pq }}{2 p}
 \\
 & = \frac{1-\sqrt{1-4 q + 4 q ^2}}{2 p}
 \\
 & = \frac{1-|1-2 q |}{2 p}
 \\
 & = \left\{
 \begin{array}{ll}
\displaystyle   \frac{q }{p } & p \geq 1/2,
   \medskip
   \\ 
   1& p \leq 1/2.
   \end{array} \right. 
 \end{align*}
 Finally,  
 by Lagrange inversion,
 see e.g. Theorem~2.10 in \cite{drmota}, 
 and the binomial theorem, we have 
\begin{align*}%
 {\P\big( S^{\scaleto{\neq 0}{6pt}}_\infty = n \big)}
   &= \frac{1}{n !} G^{(n)}(0)
  \\
  &
  = \frac{1}{n!} \frac{\partial^{n-1}}{\partial w^{n-1}} (g(w))^n_{\big| w=0}
  \\
  &
  = \frac{1}{n!} \frac{\partial^{n-1}}{\partial w^{n-1}}
  \sum_{k=0}^n \binom{n}{k} q^{n-k} p^k w^{2k}_{\big| w=0}
  \\
  &= \frac{1}{n!} \sum_{k=\lceil (n-1)/2 \rceil}^n \binom{n}{k}
  \frac{q^{n-k} p^k (2k)!}{(2k-n+1)!} w^{2k-n+1}_{\big| w=0}, 
\end{align*}
 from which \eqref{recursion-total-sol-discrete} follows. 
\end{Proofy}
\begin{Proofy} {\em of Theorem~\ref{const-weight-dc}.}
 {Recall that $X^{(k)}$ denotes
 the count of types equal to $k \geq 1$ 
 in ${\cal T}$, excluding the initial node, with 
 $X^{(k)} = 0 \mbox{ for } k > S^{\scaleto{\neq 0}{6pt}}_\infty$.}
 In what follows, we let 
\begin{align} 
  \label{def-pj-dc}
    p_j ( m_1,\ldots , m_n )
& := \P_j \big( X^{(1)} = m_1,\ldots , X^{(n)}=m_n,
 \ S^{\scaleto{\neq 0}{6pt}}_\infty = m_1+\cdots + m_n \big) 
\\ 
\nonumber
& = \P_j \big( X^{(1)} = m_1,\ldots , X^{(n)}=m_n,
 \ X^{(i)}=0 \mbox{ for all } i \geq n+1 \big), \quad j\geq 0.
\end{align}
Our proof proceeds by induction on the value of $m_1+\cdots +m_n$,
noting that when $m_1=\cdots = m_n=0$, we have
$p_j(0,\ldots, 0)=
 {\P \big( S^{\scaleto{\neq 0}{6pt}}_\infty = 0\big) = q}$.

\medskip

\noindent
  $(i)$
From the branching mechanism defining the
random tree ${\cal T}$, we have %
\begin{align}
\label{eqn-5-dc}
   p_0 ( m_1,\ldots , m_n )
    & = p \ind_{\{m_1 > m_2\}} p_0 (
   m_1-1, m_2, \ldots , m_n )
   p_1 (
{0},
\ldots
,
0
)
    \\
    \nonumber
    & \quad
    + p \sum_{\begin{subarray}{c}
      m'_i + m''_i = m_i - \ind_{\{1\leq i\leq 2\}}, \ \! 1\leq i\leq n \\
      0\leq m'_i\leq m'_{i-1}, \ \! 2\leq i\leq n \\
      0\leq m''_i\leq m''_{i-1}, \ \! 2\leq i\leq n, i\ne 3 \\
      0\leq m''_3 \leq m''_2+1
\end{subarray}} p_0\left(
    m'_1,\ldots , m'_n \right)
    p_1 \left(
m''_1,m''_2+1,m''_3,\ldots , m''_n \right),
\end{align} 
 and, {for $1 \leq j < n-1$},
\begin{align}
\label{eqn-4-dc}
 & p_j (
    m_1,\ldots , m_j,m_{j+1}+1, m_{j+2},\ldots , m_n )
    \\
    \nonumber
    & \quad
     = p \ind_{\{m_{j+1} \geq m_{j+2}\}}
 p_0 (m_1,\ldots , m_n ) p_{j+1} {(
0,
\ldots
,
0
) }
 \\
 \nonumber
 & \quad \quad 
  + p %
  \sum_{\begin{subarray}{c}
      m'_i + m''_i = m_i - \ind_{\{1\leq i=j+2\}}, \ \! 1\leq i\leq n \\
      0\leq m'_i\leq m'_{i-1}, \ \! 2\leq i\leq n \\
      0\leq m''_i\leq m''_{i-1}, \ \! 2\leq i\leq n, i\ne j+3 \\
      0\leq m''_{j+3} \leq m''_{j+2}+1
  \end{subarray}}
  p_0 ( m'_1,\ldots , m'_n )
  \\
  \nonumber
  & \qquad \qquad \qquad \qquad \qquad \qquad
  \times p_{j+1}
  ( m''_1,\ldots , {m''_{j+1}},m''_{j+2}+1,m''_{j+3}, \ldots , m''_n){,} 
\end{align}
 {while for $j=n-1$ we {have} %
\begin{align}
  \label{eqn-4-dc-2}
  p_{n-1} (
 m_1,\ldots , m_{n-1},m_n+1 )
 = {p}
  p_0 (m_1,\ldots , m_n ) p_n ( 0, \ldots , 0). 
\end{align}
 }
 We apply \eqref{eqn-4-dc} with $j=1$
 to \eqref{eqn-5-dc} to get, since
 {$p_j(
0,
\ldots
, 0
)
 = q$},
\begin{align}
\nonumber 
   & p_0 ( m_1,m_2\ldots , m_n )
     = pq \ind_{\{m_1 > m_2\}} p_0 (
   m_1-1, m_2, \ldots , m_n )
    \\
    \nonumber
    & 
    + p^2q \sum_{\begin{subarray}{c}
      m^1_i + m^2_i = m_i - \ind_{\{1\leq i\leq 2\}}, \ \! 1\leq i\leq n \\
      0\leq m^1_i\leq m^1_{i-1}, \ \! 2\leq i\leq n \\
      0\leq m^2_i\leq m^2_{i-1}, \ \! 2\leq i\leq n, i\ne 3 \\
      0\leq m^2_3 \leq m^2_2+1
\end{subarray}} p_0\left(
    m^1_1,\ldots , m^1_n \right)
 \ind_{\{m^2_2 \geq m^2_3 \}}
 p_0 (m^2_1,\ldots , m^2_n ) 
    \\
    \nonumber
    & 
    + p^2
\sum_{\begin{subarray}{c}
      m^1_i + m^2_i + m^3_i = m_i - \ind_{\{1\leq i\leq 3\}}, \ \! 1\leq i\leq n \\
      0\leq m^1_i\leq m^1_{i-1}, \ 2\leq i\leq n \\
      0\leq m^2_i\leq m^2_{i-1}, \ 2\leq i\leq n \\
      0\leq m^3_i\leq m^3_{i-1}, \ 2\leq i\leq n, i\ne 4 \\
      0\leq m^3_4 \leq m^3_3+1
\end{subarray}}
 p_0 ( m^1_1,\ldots , m^1_n )
 p_0 ( m^2_1,\ldots , m^2_n )
 p_{{2}} ( {m^3_1, m^3_2},m^3_3+1,m^3_4, \ldots , m^3_n)
. 
\end{align} 
 By repeated application of 
 \eqref{eqn-4-dc} with $j=2,\ldots,n-{2}$
 {as well as \eqref{eqn-4-dc-2}
 and using the fact that
 $m_1^k+\cdots + m_n^k\leq m-l$ for all $k=1,\ldots , n$,}
   we obtain 
$$ 
 p_0 ( m_1,\ldots ,m_n )
 = q \sum_{l=1}^n \ind_{\{m_l > m_{l+1}\}} p^l \sum_{\begin{subarray}{c}
      \sum_{k=1}^l m_i^k = m_i -\ind_{\{1\leq i\leq l\}}, \ \! 1\leq i\leq n \\
      0\leq m_i^k \leq m_{i-1}^k, \ \! 2\leq i\leq n, \ \! 1\leq k\leq l
 \end{subarray}} \prod_{k=1}^l 
 p_0 ( m_1^k,\ldots , m_n^k ).
$$ 
 Next, by the recurrence
 assumption \eqref{recursion-1-sol-j-0}
 and Proposition~\ref{radius-conv-2}, we have 
\begin{align*} 
  p_0 (m_1^k,\ldots , m_n^k ) & =
 \frac{1}{C_m} 
 b_0\big( m^k_1,\ldots,m^k_n \big)
 \P\big( S^{\scaleto{\neq 0}{6pt}}_\infty = m_1^k + \cdots + m_n^k \big)
 \\
 & =
b_0\big( m^k_1,\ldots,m^k_n \big)
 q^{1 + m^k_1 + \cdots + m^k_n } p^{m^k_1 + \cdots + m^k_n}
,
\end{align*} 
 hence 
\begin{align*}
 & p_0 ( m_1,\ldots , m_n ) \\
   & \quad = q \sum_{l=1}^n \ind_{\{m_l > m_{l+1}\}} p^l \sum_{\begin{subarray}{c}
      \sum_{k=1}^l m_i^k = m_i -\ind_{\{1\leq i\leq l\}}, \ \! 1\leq i\leq n \\
      0\leq m_i^k \leq m_{i-1}^k, \ \! 2\leq i\leq n, \ \! 1\leq k\leq l
  \end{subarray}} \prod_{k=1}^l b_0 \big( m^k_1,\ldots,m^k_n \big)
  q^{1+ m^k_1 + \cdots + m^k_n } p^{m^k_1 + \cdots + m^k_n} \\
   & \quad = q (pq)^{m_1 + \cdots + m_n} \sum_{l=1}^n \ind_{\{m_l > m_{l+1}\}} \sum_{\begin{subarray}{c}
      \sum_{k=1}^l m_i^k = m_i -\ind_{\{1\leq i\leq l\}}, \ \! 1\leq i\leq n \\
      0\leq m_i^k \leq m_{i-1}^k, \ \! 2\leq i\leq n, \ \! 1\leq k\leq l
    \end{subarray}} \prod_{k=1}^l b_0\left( m_1^k,\ldots,m_n^k \right), 
\end{align*}
which shows
 \eqref{recursion-1-sol-j-0}
 for $j=0$ from
 \eqref{recursion-total-sol-discrete}
 and the recursive definition
 \eqref{recursion-1-coef-general-discrete-j}
  of $b_0$.

\noindent
$(ii)$ We iterate \eqref{eqn-4-dc} over
{$n-j-1$} steps
 {and then use
 \eqref{eqn-4-dc-2} to obtain} 
\begin{align*}
  & p_j ( m_1,\ldots ,
  {m_j} , m_{j+1} + 1 ,
  {m_{j+2}}, \ldots , m_n )
  \\
  & = q
  \sum_{l=1}^{n-j} \ind_{\{m_{j+l} - \ind_{\{l\geq 2\}} \geq m_{j+l+1}\}} p^l
  \sum_{\begin{subarray}{c}
      \sum_{k=1}^l m_i^k = m_i -\ind_{\{j+2\leq i\leq j+l\}}, \ \! 1\leq i\leq n \\
      0\leq m_i^k \leq m_{i-1}^k, \ \! 2\leq i\leq n, \ \! 1\leq k\leq l
  \end{subarray}} \prod_{k=1}^l p_0
  ( m_1^k, \ldots , m_n^k ) \\
  & = q (pq)^{1 + m_1 + \cdots + m_n} \sum_{l=1}^{n-j} \ind_{\{m_{j+l} - \ind_{\{l\geq 2\}} \geq m_{j+l+1}\}}
  \hskip-.6cm
  \sum_{\begin{subarray}{c}
      \sum_{k=1}^l m_i^k = m_i -\ind_{\{j+2\leq i\leq j+l\}}, \ \! 1\leq i\leq n \\
      0\leq m_i^k \leq m_{i-1}^k, \ \! 2\leq i\leq n, \ \! 1\leq k\leq l
    \end{subarray}} \prod_{k=1}^l b_0\big(m^k_1,\ldots,m^k_n\big), 
\end{align*}
which shows
 \eqref{recursion-1-sol-j-0}
 for $j\geq 1$ from
 \eqref{recursion-total-sol-discrete}
 and 
 \eqref{recursion-1-coef-general-discrete-j}.
\end{Proofy}
\begin{Proofy} {\em of Corollary~\ref{fjklds13-0}}.   
\noindent
 Let 
  \begin{equation}
    \label{eqn-22}
      B_j^\sigma(m)
 := 
 C_m
 \E_j \left[
            \prod_{k=1}^{m { + j}} 
   \sigma (k)^{X^{(k)}}
   \ \! \Big| \ \!
        S^{\scaleto{\neq 0}{6pt}}_\infty = m
        \right]
 ,
 \quad
 j \geq 0,
\end{equation} 
 with $B_j^\sigma (0)= 1$.
 By Theorem~\ref{const-weight-dc},
 we have
$$ B_j^\sigma(m)
 =
 \sum_{n=1}^{m{+j}}
 \sum_{\begin{subarray}{c}
    (m_1,\ldots,m_n)\in {\mathbb K}_{j,n} \\
    m_1 + \cdots + m_n = m
 \end{subarray}}
 b_j^\sigma (m_1,\ldots,m_n)
 ,
 $$ 
 where $$
 b_j^\sigma (m_1,\ldots,m_n)
 :=
 b_j (m_1,\ldots,m_n)
 \prod_{k=1}^n \sigma(k)^{m_k}. 
$$
 By the induction relation \eqref{recursion-1-coef-general-discrete-j},
 i.e. 
$$
 b_j^\sigma (m_1,\ldots,m_n) 
 =
 \sum_{l=1}^{n-j} \ind_{\{m_{j+l} > m_{j+l+1}\}}
    \hskip-0.4cm
    \sum_{\begin{subarray}{c}
      \sum_{k=1}^l m_i^k = m_i -\ind_{\{j < i\leq j+l\}}, \ \! 1\leq i\leq n \\
      0\leq m_i^k \leq m_{i-1}^k, \ \! 2\leq i\leq n, \ \! 1\leq k\leq l
    \end{subarray}} \prod_{k=1}^l b_0^\sigma\left( m_1^k,\ldots,m_n^k \right) 
$$ 
 we have 
\begin{align}
\nonumber %
 & B_j^\sigma (m+1)
 = \sum_{\begin{subarray}{c} m_1+\cdots +m_n = m+1, \ n\geq 1, \\ 1\leq m_i \leq m_{i-1}, \ 2\leq i\leq n \end{subarray}} b^\sigma_j (m_1,\ldots,m_n)
    \\
    \nonumber
    & \quad = \sum_{n=j+1}^{m+j+1}\sum_{\begin{subarray}{c}
      m_1+\cdots +m_n = m+1 \\
      1\leq m_i \leq m_{i-1}, \ 2\leq i\leq n
    \end{subarray}} \sum_{l=1}^{n-j}
       \ind_{\{
        m_{j+l} > m_{j+l+1}
        \}}
    \sum_{\begin{subarray}{c}
      \sum_{k=1}^l m^k_i = m_i -\ind_{\{j < i\leq j+l\}}, \ \! 1\leq i\leq n \\
      0\leq m^k_i \leq m_{i-1}^k, \ 2\leq i\leq n, \ \! 1\leq k\leq l
    \end{subarray}} \prod_{k=1}^l
    b^\sigma_0 ( m^k_1,\ldots,m^k_n )
    \\
\nonumber
&\quad = \sum_{l=1}^{m+1}
\sum_{n'=1}^{m+1-l} \sum_{\begin{subarray}{c}
      m'_1+\cdots +m'_{n'} = m+1-l \\
      1\leq m'_i \leq m'_{i-1}, \ 2\leq i\leq n'
\end{subarray}}
\sum_{\begin{subarray}{c}
      \sum_{k=1}^l m^k_i = m'_i, \ \! 1\leq i\leq n' \\
      0\leq m^k_i \leq m_{i-1}^k, \ 2\leq i\leq n', \ \! 1\leq k\leq l
\end{subarray}} \prod_{k=1}^l
b^\sigma_0 ( m^k_1,\ldots,m^k_{n'} )
\\
\nonumber
&\quad = \sum_{l=1}^{m+1}
 \sum_{\begin{subarray}{c}
      m_1+\cdots + m_l = m+1-l \\
      m_1,\ldots , m_l \geq 0
 \end{subarray}}
 ~
 \sum_{n'\geq 1}\sum_{\begin{subarray}{c}
      m^k_1+\cdots +m^k_{n'} = m_k, \ \! 1\leq k\leq l \\
      0\leq m^k_i \leq m_{i-1}^k, \ 2\leq i\leq n', \ \! 1\leq k\leq l \\
      \text{at least one of } m^k_{n'}, \ \! 1\leq k\leq l \text{ is nonzero}
\end{subarray}} \prod_{k=1}^l
b^\sigma_0( m^k_1,\ldots,m^k_{n'} )
\\
\nonumber
&\quad = \sum_{l=1}^{m+1}
 \sum_{\begin{subarray}{c}
      m_1 + \cdots + m_l = m+1-l \\
     m_1, \ldots , m_l \geq 0
\end{subarray}} \prod_{k=1}^l
 \sum_{n_k \geq 0} \sum_{\begin{subarray}{c}
      m^k_1+\cdots + m^k_{n_k} = m_k \\
      1\leq m^k_i \leq m_{i-1}^k, \ 2\leq i\leq n_k
    \end{subarray}} b^\sigma_0 ( m^k_1,\ldots,m^k_{n_k} )
\\
\label{fjkld3-0}
&\quad = \sum_{l=1}^{m+1}
\left( \prod_{k=j+1}^{j+l} \sigma(k) \right)
\sum_{\begin{subarray}{c}
      m_1 + \cdots + m_l = m+1 \\
      m_1\geq 1, \ldots , m_l \geq 1
\end{subarray}} \prod_{k=1}^l B_0^\sigma (m_k-1)
    ,
 \qquad
  m\geq 0,
\end{align} 
 where in the third equality we made the change of variables
 $m'_i = m_i - \ind_{\{j < i\leq j+l\}}$.
 Let now 
\begin{align*} 
D^{(k)}_j(m) & :=
C_m \E_j \big[ X^{(k)} \ \! \big| \ \! S^{\scaleto{\neq 0}{6pt}}_\infty = m \big] 
\\
 & =
 \sum_{n=\max ( k  ,j)}^{m+j}
  \sum_{\begin{subarray}{c}
    (m_1,\ldots,m_n)\in {\mathbb K}_{j,n} \\
    m_1 + \cdots + m_n = m
    \end{subarray}}
  m_k b_j(m_1,\ldots,m_n)
  \\
   & =
 \frac{\pt}{\pt \sigma (k)}\bigg|_{\mathbf \sigma = \mathbf 1}
 B_j^\sigma (m)
,
 \quad l =1,\ldots , m + j,
 \ 
 \quad j,m\geq 0, 
\end{align*} 
 with initial values $D^{(k)}_j(0) = 0$.
 By \eqref{fjkld3-0}, for $m\geq 0$ we have 
\begin{align*}
  &  D^{(k)}_j(m+1) = \frac{\pt}{\pt \sigma (k)}\bigg|_{\mathbf \sigma = \mathbf 1} [x^{m+1}] \sum_{l=1}^\infty \left( \prod_{k'=j+1}^{j+l} \sigma (k') \right) \left( \sum_{n=1}^\infty
    B_0^\sigma (n-1)
    x^n \right)^l \\
    & \quad
    = [x^{m+1}] \sum_{l=1}^\infty \ind_{\{ j < k \leq j+l \}} \left( \prod_{k'=j+1, k' \ne k}^{j+l} \sigma ( k') \right) \left( \sum_{n=1}^\infty
    B_0^\sigma (n-1)
    x^n \right)^l \bigg|_{\mathbf \sigma = \mathbf 1} \\
    &  \quad \quad + [x^{m+1}] \sum_{l=1}^\infty \left( \prod_{k'=j+1}^{j+l}
    \sigma ( k')
    \right) l \left( \sum_{n=1}^\infty
    B_0^\sigma (n-1)
    x^n \right)^{l-1} \left( \sum_{n=1}^\infty \frac{\pt}{\pt \sigma ( k)}
    B_0^\sigma (n-1)
    x^n \right) \bigg|_{\mathbf \sigma = \mathbf 1} \\
    &  \quad = \ind_{\{ j < k \}} [x^{m+1}] \sum_{l=k-j}^\infty \left( \sum_{n=1}^\infty
    B_0^{\bf 1} (n-1)
    x^n \right)^l \\
    &  \quad \quad + [x^{m+1}] \sum_{l=1}^\infty l \left( \sum_{n=1}^\infty
    B_0^{\bf 1} (n-1)
    x^n \right)^{l-1} \left( \sum_{n=1}^\infty D^{(0)}_k(n) x^{n+1} \right), 
\end{align*}
 where $[x^{m+1}]$ is the operator extracting the coefficient of the
 term $x^{m+1}$ from the series following it. 
 Thus,
\begin{align*}
  \sum_{m=0}^\infty D^{(k)}_j(m+1) x^{m+1}
  & = \ind_{\{ j < k \}} \sum_{l=k-j}^\infty \left( \sum_{n=1}^\infty
  B_0^{\bf 1} (n-1)
  x^n \right)^l
  \\
   & \quad + \sum_{l=1}^\infty l \left( \sum_{n=1}^\infty
  B_0^{\bf 1} (n-1)
  x^n \right)^{l-1} \left( \sum_{n=1}^\infty D^{(0)}_k(n-1) x^n \right).
\end{align*}
 By \eqref{eqn-22} and Proposition~\ref{radius-conv-2}, 
 we have 
\begin{equation*}
  \sum_{n=1}^\infty
  B_0^{\bf 1} (n-1)
  x^n = \sum_{n=1}^\infty C_{n-1} x^n = \frac{1-\sqrt{1-4 x}}{2},
\end{equation*}
which implies
\begin{align*}
  \sum_{l=k}^\infty \left( \sum_{n=1}^\infty
  B_0^{\bf 1} (n-1)
  x^n \right)^l
  = x^k \left( \frac{1-\sqrt{1-4 x}}{2x} \right)^{k+1}, 
\end{align*}
and
\begin{equation*}
  \sum_{l=1}^\infty l \left( \sum_{n=1}^\infty
  B_0^{\bf 1} (n-1)
  x^n \right)^{l-1} =
  \left( \frac{1-\sqrt{1-4 x}}{2x} \right)^2. 
\end{equation*}
 Hence, the unconditional expected value of $X^{(k)}$ is given by 
\begin{align*} 
\E_j \big[ X^{(k)} \big]
 & = 
  \sum_{m={0} }^\infty
  \E_j \big[ X^{(k)} \ \! \big| \ \! S^{\scaleto{\neq 0}{6pt}}_\infty = m \big]
 \P \big( S^{\scaleto{\neq 0}{6pt}}_\infty = m\big) 
 \\
 &=
 q \sum_{m=0}^\infty D^{(k)}_j(m+1 ) (pq)^{m+1}
 \\
    &= \ind_{\{ j < k \}} \frac{1}{p} \left( \frac{1-\sqrt{1-4 pq}}{2} \right)^{k+1-j}
    +
    \frac{1}{pq}
    \left(
    \frac{1-\sqrt{1-4 pq}}{2}
    \right)^2 
    \E_0 \big[ X^{(k)} \big]
    \\
    &=
    \ind_{\{ j < k \}} 
    p^{k-j}
    +
    \frac{p}{q}
    \E_0 \big[ X^{(k)} \big].
\end{align*}
 When $j=0$, this yields 
\begin{equation*}
  \E_0 \big[ X^{(k)} \big]
  = \frac{q}{\sqrt{1-4 pq}} \left( \frac{1-\sqrt{1-4 pq}}{2} \right)^k
 = \frac{qp^k}{q-p}
  , 
\end{equation*}
and in general
 we obtain 
\begin{align*} 
  \E_j \big[ X^{(k)} \big]
  & = \frac{1}{p}
  \ind_{\{ j < k\}}
  \left( \frac{1-\sqrt{1-4 pq}}{2} \right)^{k+1-j}
  +
 \frac{1}{p\sqrt{1-4 pq}} \left( \frac{1-\sqrt{1-4 pq}}{2} \right)^{k+2}
 \\
  & =
 \ind_{\{ j < k\}}
  p^{k-j}
  +
  \frac{p^{k+1}}{q-p}
  . 
\end{align*} 
 Hence, when $j=0$ we have
\begin{equation*}
  \E_0 \big[ X^{(k)} \big]
  = q \sum_{n=k}^\infty \binom{2n-k}{n} (pq)^n, 
\end{equation*}
 and in general we obtain 
\begin{equation*}
  \begin{split}
    \E_j \big[ X^{(k)} \big] 
    &=
    q   \ind_{\{ j < k\}}
    \sum_{n=k-j}^\infty 
    \frac{k+1-j}{n+1} \binom{2n-k+j}{n} 
    (pq)^n
 + 
 q \sum_{n=k}^\infty
 \binom{2n-k}{n+1} 
 (pq)^n
, 
\end{split}
\end{equation*}
 which yields \eqref{fjkld3-1}.
\end{Proofy}
\begin{Proofy} {\em of Corollary~\ref{fjklds13-1}}.   
\noindent
 Using \eqref{fjkld3-1}, we have 
\begin{align*} 
 &   \E_j \left[ \frac{X^{(k)}}{
        S^{\scaleto{\neq 0}{6pt}}_\infty }
      \ \! \Big| \ \!
      S^{\scaleto{\neq 0}{6pt}}_\infty \geq 1
      \right] 
 = 
 \frac{1}{p} \sum_{m=1}^\infty
 \frac{1}{m}
 \E_j \big[ X^{{(k)}} \ \! \big| \ \! S^{\scaleto{\neq 0}{6pt}}_\infty = m \big]
 \P \big( S^{\scaleto{\neq 0}{6pt}}_\infty = m \big)
  \\
  &
   =
   \frac{q}{p}
   \ind_{\{ j < k\}}
    \sum_{m=k-j}^\infty 
    \frac{k+1-j}{m+1} \binom{2m-k+j}{m} 
    \frac{(pq)^m}{m}
 + 
 \frac{q}{p}
 \sum_{m=k}^\infty
 \binom{2m-k}{m+1} 
 \frac{(pq)^m}{m}
  \\
  &
   =
   \frac{q}{p}
   \ind_{\{ j < k\}}
   \int_0^{pq}
   \sum_{m=k-j}^\infty 
    \frac{k+1-j}{m+1} \binom{2m-k+j}{m} 
    x^{m-1}
    dx
    + 
 \frac{q}{p}
 \int_0^{pq}
 \sum_{m=k}^\infty
 \binom{2m-k}{m+1} 
 x^{m-1} dx 
  \\
  &
   =
   \frac{q}{p}
   \ind_{\{ j < k\}}
   \int_0^{pq}
   \frac{1}{x^2}
   \left( \frac{1-\sqrt{1-4 x}}{2} \right)^{k+1-j}
   dx 
 + 
 \frac{q}{p}
 \int_0^{pq}
 \frac{1}{x^2\sqrt{1-4 x}} \left( \frac{1-\sqrt{1-4 x}}{2} \right)^{k+2}
 dx
 \\
  &
   =
   \frac{q}{p}
   \ind_{\{ j < k\}}
   \left(
   (k + 1 - j ) B\left( \frac{1 - \sqrt{1 - 4 p q}}{2} ; k-j; 0\right) 
 - \frac{1}{p q}
  \left(\frac{1 - \sqrt{1 - 4 p q}}{2} )^{k+1-j} \right)\right)
   \\
   &
   \quad
 +     \frac{q}{p}
 B \left( \frac{1 - \sqrt{1 - 4 p q}}{2} ; 1 + k, -1 \right)
.
\end{align*}

\end{Proofy}
\begin{Proofy} {\em of Proposition~\ref{B-special}}.   
 Taking $j=0$ and 
$$
\sigma(k) := 1+\frac{\gamma }{k},
\quad k\geq 1,
$$
 in \eqref{fjkld3-0} and denoting $B_j^\sigma$ by $B_j^\gamma$,
 we have 
 \begin{equation}
\nonumber %
 B_0^\gamma (n+1) = \sum_{l=1}^{n+1} \binom{l+\gamma}{l} \sum_{\begin{subarray}{c}
      m_1 + \cdots + m_l = n+1 \\
      m_1, \ldots , m_l \geq 1
    \end{subarray}} \prod_{k=1}^l B_0^\gamma (m_k-1), 
  \end{equation}
  and by the Fa\`a di Bruno formula
  in Lemma~\ref{lemma-comb-generalized} below 
  we find that $B_0^\gamma (n)$ is the coefficient of $x^{n}$ in the series
$$ 
  \sum_{l=1}^\infty \binom{l+\gamma}{l}
  \left( \sum_{n=1}^\infty \binom{(2+\gamma)n-2}{n-1}
  \frac{x^n}{n} \right)^l.
$$
 By Lemma~\ref{A2} below,
 denoting by $\Phi^{-1}_\gamma$ the inverse function of 
 \begin{equation}
\nonumber %
  \Phi_\gamma (w) := w(1-w)^{1+\gamma}, \quad w\in\mathbb C, 
\end{equation}
 we have 
\begin{align}
 \nonumber %
  \sum_{l=1}^\infty \binom{l+\gamma}{l}
  \left( \sum_{n=1}^\infty \binom{(2+\gamma)n-2}{n-1}
  \frac{x^n}{n} \right)^l
&  =
  \sum_{l=1}^\infty \binom{l+\gamma}{l}
  \left(
  \sum_{n=1}^\infty F_{{n-1}} (\gamma + 2,\gamma +1)
  x^n \right)^l
\\
  \nonumber
  &  =
  \sum_{l=1}^\infty \binom{l+\gamma}{l}
  \big( \Phi^{-1}_\gamma (x) \big)^l
    \\
  \nonumber
  &  =
  \big( 1 - \Phi^{-1}_\gamma (x) \big)^{-\gamma -1} {-1}
    \\
  \nonumber
  &  =
  \frac{1}{x} \Phi^{-1}_\gamma (x) -1
    \\
  \nonumber
  & =
  \sum_{n=0}^\infty F_n(\gamma + 2,\gamma +1)
  x^n
  , 
\end{align}
  which yields \eqref{hhfdsjf}.
\end{Proofy}
\noindent
 We also recall the following 
 version of the Fa\`a di Bruno formula
 which is used in
 the proofs of Propositions~\ref{B-special}
 and \ref{hjkfd21},
 see for example Theorem~5.1.4 in \cite{stanley}.
\begin{lemma}
\label{lemma-comb-generalized}
For any two sequences $(\alpha_n )_{n\geq 1}$, 
$(\beta_n)_{n\geq 1}$, the coefficient of $x^m$, $m\geq 1$,
in the series
\begin{equation*}
  \sum_{l=1}^\infty \alpha_l \bigg( \sum_{n=1}^\infty \beta_n x^n \bigg)^l
\end{equation*}
is given by
\begin{equation*}
    \sum_{l=1}^m \alpha_l \sum_{\begin{subarray}{c}
       m_1 + \cdots + m_l = m \\
      m_1 , \ldots , m_l \geq 1
    \end{subarray}} \beta_{m_1}\cdots \beta_{m_l}. 
\end{equation*}
\end{lemma} 
\noindent
 The following lemma was used in the proof of Proposition~\ref{B-special}. 
\begin{lemma} 
  \label{A2}
  The inverse function
  $\Phi_\gamma^{-1}$
  of 
\begin{equation}
  \Phi_\gamma (w) := w(1-w)^{1+\gamma}, \quad w\in\mathbb C, 
\end{equation}
 admits the expansion 
\begin{equation*}
  \Phi^{-1}_\gamma (x) = 
  \sum_{n=1}^\infty F_{n-1} (\gamma + 2 , \gamma +1) x^n. 
\end{equation*}
\end{lemma}
\begin{Proof}
 Since $\Phi_\gamma $ is analytic near $w=0$ and $\Phi_\gamma (0)=0$, $\Phi'_\gamma (0) = 1\ne 0$, by the Lagrange inversion theorem, the inverse function of $\Phi_\gamma $ is given by the power series
\begin{equation}
 \nonumber %
\Phi^{-1}_\gamma (z) = \sum_{n=1}^\infty \frac{\alpha_n}{n!} z^n,
\end{equation} 
 where
\begin{align*}
  \alpha_n &= \lim_{w\to0} \frac{\partial^{n-1}}{\partial w^{n-1}} \left( \frac{w}{\Phi_\gamma (w)} \right)^n
  \\
  &
  = \lim_{w\to0} \frac{\partial^{n-1}}{\partial w^{n-1}} (1-w)^{-(1+\gamma) n}
  \\
    &= \lim_{w\to0} \frac{\partial^{n-1}}{\partial w^{n-1}}
    \sum_{k=0}^\infty \binom{k+(1+\gamma) n-1}{k} w^k
    \\
    &
    = (n-1)! \binom{(2+\gamma) n-2}{n-1}.
\end{align*}
\end{Proof}
\section{Proofs - continuous-time setting} 
\label{s5}
\begin{Proofy} {\em of Proposition~\ref{progeny}}. 
\noindent
 We denote by 
$$
 \widebar{F}_\rho(t) := \P\left( T_\varnothing>t\right) = \int_t^\infty \rho (r) dr 
 , \qquad t \geq 0, 
$$
 the tail cumulative distribution function of $\rho$,
 and let 
 $p_t(n) := \P( N_t = n )$,
 $n\geq 0$, with 
 \begin{equation}
   \nonumber %
   p_t( {0} ) = \P( N_t = {0} ) =
   \P\left( T_\varnothing>t\right) = \widebar{F}_\rho(t),
  \quad t\in \real_+.\end{equation}
For $n\geq {1}$, by the
relation $\{ N_t \geq 1\} \subset \{ T_\varnothing\leq t \}$
 and independence of branches, 
 denoting by
 $( N^1_t)_{t\in \real_+}$
 and
 $( N^2_t)_{t\in \real_+}$
 two independent copies of
 $( N_t)_{t\in \real_+}$, we have 
\begin{align} 
\nonumber %
  p_t(n) &= \P( N_t = n )
  \\
  \nonumber
  &= \E \big[ \P ( N_t = n, T_\varnothing\leq t
     \mid T_\varnothing ) \big] \\
  \nonumber
  &= \E \big[ \P( N^1_s + N^2_s = n-1 )_{| {s=t-T_\varnothing}} \ind_{\{ T_\varnothing\leq t \}} \big]
  \\
  \nonumber
  &= \E \big[ p_s^{*2}(n-1)_{|{s=t- T_\varnothing}} \ind_{\{ T_\varnothing\leq t \}} \big]
  \\
  \nonumber
  &= \int_0^t
  {\rho (t-s)} p_s^{*2}(n-1)
  ds,
\end{align} 
 where $*$ is the discrete convolution product.    
 As the distribution $\rho$
 is exponential with parameter $\lambda$,
 we have 
  \begin{equation}
    \label{recursion-total}
p_t(n) =
  \begin{cases}
    \displaystyle e^{-\lambda t}, & n={0},
    \\
    \displaystyle
    \lambda \int_0^t e^{-(t-s)\lambda} p_s^{*2}(n-1) ds
    = \lambda \int_0^t e^{-(t-s)\lambda }
    \displaystyle \sum_{\begin{subarray}{c}
    n_1 + n_2 = n-1 \\
    n_1, n_2 \geq 0
     \end{subarray}} p_s(n_1) p_s(n_2) ds
     , & n\geq {1}. 
  \end{cases}
\end{equation}
Multiplying both sides of the third equality in \eqref{recursion-total} by
 $z^n$ and summing over $n\geq {1}$ gives 
\begin{equation*}
 G_t(z) - z e^{-\lambda t} = z \lambda \int_0^t e^{-(t-s)\lambda} G_s(z)^2 ds, 
\end{equation*}
 which in turns yields the Bernoulli ODE
 \begin{equation}
   \label{Bernoulli}
    \frac{d}{dt} G_t(z) + \lambda G_t(z) = \lambda z G_t(z)^2, \quad t>0, 
\end{equation}
 with initial condition $G_0(z) = z$ 
since $p_0(n) = \ind_{\{n= {0} \}}$. 
The solution of \eqref{Bernoulli} is then obtained by 
a standard argument, which allows us to conclude to \eqref{pgf-ct}. 
\end{Proofy}
\begin{Proofy} {\em of Theorem~\ref{const-l-weight}}. 
 {Recall that $X_t^{(i)}$ denotes the count of types equal to $i \geq 1$ until time~$t$, excluding the initial node. 
 Similarly to \eqref{def-pj-dc},}
 we let 
\begin{align*}
 p_{t,j} (m_1,\ldots,m_n)
 & :=
\P_j \big( X_t^{(1)} = m_1,\ldots , X_t^{(n)}=m_n,
 \ %
N_t = m_1 + \cdots + m_n
\big)
\\
 & = {
\P_j \big( X_t^{(1)} = m_1,\ldots , X_t^{(n)}=m_n, \ X_t^{(i)}=m_n 
\mbox{~for~all~}
i \geq n+1 \big)}, 
\end{align*}
 {$j \geq 0$}. 
Our proof proceeds by induction on the value of $m_1+\cdots +m_n$,
with 
$$
p_{t,j}(0,\ldots, 0)={\P ( N_t = 0) = e^{-\lambda t}}
$$
when $m_1=\cdots = m_n=0$. 
 
\medskip

 We note that the branching chain
  $( X_t)_{t\geq 0}$ with initial type $0$ has $m_i$ branches with type $i$ for each $i\geq 1$, then it must have
$\left( 1 + m_1+\cdots +m_n \right)$ branches with type $0$, since each branch with type $0$, except the initial one, has one and only one brother with a positive type.   

  \noindent 
$(i)$ For $j=0$, we have 
\begin{align}
  \label{eqn-5}
  &\ p_{t,0} ( m_1,\ldots , m_n ) 
 = \ind_{\{
    m_1 > m_2
    \}} \lambda \int_0^te^{-(t-s)\lambda} p_{s,0}
  ( m_1-1 ,m_2,\ldots , m_n) p_{s,1}(1) ds
  \\
  \nonumber
    & \quad + \lambda \int_0^t e^{-(t-s)\lambda} \sum_{\begin{subarray}{c}
      m'_i + m''_i = m_i - \ind_{\{1\leq i\leq 2\}}, \ \! 1\leq i\leq n \\
      0\leq m'_i\leq m'_{i-1}, \ 2\leq i\leq n \\
      0\leq m''_i\leq m''_{i-1}, \ 2\leq i\leq n, \ i\ne 3 \\
      0\leq m''_3 \leq m''_2+1
  \end{subarray}} p_{s,0}( 
 m'_1, \ldots , m'_n )
 p_{s,1} ( 
 m''_1 , m''_2+1 , m''_3, \ldots , m''_n ) 
 ds,
\end{align} 
 and, for {$1 \leq j < n-1$},
\begin{align}
\label{eqn-4}
& p_{t,j} ( m_1,\ldots , {m_j} , m_{j+1} +1 , m_{j+2}, \ldots , m_n)
 \\
  \nonumber
  &
  = \ind_{\{m_{j+1} \geq m_{j+2}\}} \lambda \int_0^t e^{-(t-s)\lambda}
  p_{s,0} ( m_1, \ldots , m_n ) {p_{s,j+1}(0,\ldots , 0)} ds
  + \lambda \int_0^t e^{-(t-s)\lambda}
  \\
  \nonumber
  &
  \sum_{\begin{subarray}{c}
      m'_i + m''_i = m_i - \ind_{\{1\leq i=j+2\}}, \ \! 1\leq i\leq n \\
      0\leq m'_i\leq m'_{i-1}, \ 2\leq i\leq n \\
      0\leq m''_i\leq m''_{i-1}, \ 2\leq i\leq n, \ i\ne j+3 \\
      0\leq m''_{j+3} \leq m''_{j+2}+1
  \end{subarray}} p_{s,0}(
       m'_1 , \ldots , m'_n )
 {p_{s , j+1}} ( 
 m''_1,\ldots , {m''_{j+1}}, m''_{j+2}+1, m''_{j+3},
 \ldots, m''_n)  ds, 
\end{align} 
 {while for $j = n-1$ we have %
\begin{align}
\label{eqn-4-2}
 p_{t,n-1} ( m_1,\ldots , m_{n-1} , m_n +1)
 =
  \lambda \int_0^t e^{-(t-s)\lambda}
  p_{s,0} ( m_1, \ldots , m_n ) p_{s,n}{(0,\ldots , 0)} ds
.
\end{align} 
} Since $p_{t,j}{(0,
\ldots
, 0
)} = e^{-\lambda t}$, we apply \eqref{eqn-4} with $j=1$ to \eqref{eqn-5} to get
\begin{align*}
  &\ p_{t,0} ( m_1 , \ldots , m_n )
  =  \ind_{\{
    m_1 > m_2
    \}} \lambda e^{-\lambda t} \int_0^t
  p_{s,0}
  ( m_1-1 , m_2 , \ldots , m_n) ds \\
  & \quad + \ind_{\{
    m_2 > m_3
    \}} \lambda^2 e^{-\lambda t} \int_0^t \int_0^s \sum_{\begin{subarray}{c}
      m^1_i + m^2_i = m_i - \ind_{\{1\leq i\leq 2\}}, \ \! 1\leq i\leq n \\
      0\leq m^1_i\leq m^1_{i-1}, \ 2\leq i\leq n \\
      0\leq m^2_i\leq m^2_{i-1}, \ 2\leq i\leq n
  \end{subarray}}
  p_{s,0} ( m^1_1 , \ldots , m^1_n ) 
  p_{r,0}( m^2_1, \ldots , m^2_n ) dr ds \\
  &\quad + \int_0^t \int_0^s \lambda^2 e^{(r-t)\lambda}
  \\
  & \quad
  \sum_{\begin{subarray}{c}
      m^1_i + m^2_i + m^3_i = m_i - \ind_{\{1\leq i\leq 3\}}, \ \! 1\leq i\leq n \\
      0\leq m^1_i\leq m^1_{i-1}, \ 2\leq i\leq n \\
      0\leq m^2_i\leq m^2_{i-1}, \ 2\leq i\leq n \\
      0\leq m^3_i\leq m^3_{i-1}, \ 2\leq i\leq n, i\ne 4 \\
      0\leq m^3_4 \leq m^3_3+1
  \end{subarray}} p_{s,0}( 
   m^1_1, \ldots , m^1_n ) 
   p_{r,0} ( m^2_1,\ldots , m^2_n )
   {p_{r,2}} ( 
  {m^3_1} , m^3_2 , m^3_3 +1, m^3_4, \ldots ,m^3_n )
  dr ds.
\end{align*}
 By repeated application of
 \eqref{eqn-4} with $j=2,\ldots,n-{2}$ {as well as
 \eqref{eqn-4-2}}, we obtain 
\begin{align}
  \label{eqn-6-1}
  &\ p_{t,0} ( m_1,\ldots , m_n) 
  \\
 \nonumber
 & 
   = e^{-\lambda t} \sum_{l=1}^n \lambda^l \ind_{\{
    m_l > m_{l+1}
    \}} 
  \int_{0\leq s_l\leq \cdots s_1\leq t}
 \sum_{\begin{subarray}{c}
      \sum_{k=1}^l m^k_i = m_i -\ind_{\{1\leq i\leq l\}}, \ \! 1\leq i\leq n \\
      0\leq m^k_i \leq m_{i-1}^k, \ 2\leq i\leq n, \ \! 1\leq k\leq l
  \end{subarray}} \prod_{k=1}^l p_{s_k,0}
  ( m^k_1, \ldots , m^k_n) ds_l \cdots ds_1
\\
 \nonumber
  & = e^{-\lambda t} \sum_{l=1}^n  \frac{
  \lambda^l}{l!} 
  \ind_{\{
      m_l > m_{l+1}
      \}}
  \sum_{\begin{subarray}{c}
      \sum_{k=1}^l m^k_i = m_i -\ind_{\{1\leq i\leq l\}}, \ \! 1\leq i\leq n \\
      0\leq m^k_i \leq m_{i-1}^k, \ 2\leq i\leq n, \ \! 1\leq k\leq l
  \end{subarray}} \prod_{k=1}^l \int_0^t p_{s,0}
  ( m^k_1 , \ldots , m^k_n) ds.
\end{align} 
Observe that in multi-index notation, the constraint in the
 above summation reads 
\begin{equation*}
  \sum_{k=1}^l (m^k_1,\ldots,m^k_n) = (m_1,\ldots,m_n) -(\underbrace{\overbrace{1,\ldots,1}^{l}, 0,\ldots,0}_{n}).
\end{equation*}
Thus, the proof can be conducted by induction over the set of multi-indices
$$
\{(m_1,\ldots,m_n) \ : \ m_1 \geq \cdots \geq m_n \geq 0 \}
$$
in the back-diagonal order.
The induction starts from the initial multi-index $\varnothing$, in which case the result follows from $a_0 (\varnothing ) = 1$ and $p_{t,0}
{(0,\ldots , 0)}
= e^{-\lambda t}$. %
 Writing the induction hypothesis as 
$$p_{s,0} ( m^k_1, \ldots , m^k_n ) 
 = a_0 ( m^k_1,\ldots,m^k_n ) e^{-\lambda s}
(1-e^{-\lambda s})^{m^k_1+\cdots +m^k_n} 
$$
 and using \eqref{eqn-6-1}, we obtain 
 \begin{align*}
  & p_{t,0} ( 
  m_1, \ldots , m_n 
 )
   \\
   & =
   e^{-\lambda t} \sum_{l=1}^n \frac{\ind_{\{
       m_l > m_{l+1}
       \}}}{l!} 
  \sum_{\begin{subarray}{c}
      \sum_{k=1}^l m^k_i = m_i -\ind_{\{1\leq i\leq l\}}, \ \! 1\leq i\leq n \\
      0\leq m^k_i \leq m_{i-1}^k, \ 2\leq i\leq n, \ \! 1\leq k\leq l
  \end{subarray}} \prod_{k=1}^l \int_0^t p_{s,0}
  ( m^k_1, \ldots , m^k_n) ds
   \\
   & =
   e^{-\lambda t} \sum_{l=1}^n \frac{\ind_{\{
       m_l > m_{l+1}
       \}}}{l!} 
   \hskip-0.6cm
   \sum_{\begin{subarray}{c}
      \sum_{k=1}^l m^k_i = m_i -\ind_{\{1\leq i\leq l\}}, \ \! 1\leq i\leq n \\
      0\leq m^k_i \leq m_{i-1}^k, \ 2\leq i\leq n, \ \! 1\leq k\leq l
  \end{subarray}} \prod_{k=1}^l \int_0^t
  a_0 ( m^k_1,\ldots,m^k_n ) e^{-\lambda s}
  (1-e^{-\lambda s})^{m^k_1+\cdots +m^k_n} ds
\\
  & = e^{-\lambda t}
  (1-e^{-\lambda t})^{m_1+\cdots +m_n} \sum_{l=1}^n \frac{\ind_{\{
      m_l > m_{l+1}
      \}}}{l!} \hskip-0.4cm
   \sum_{\begin{subarray}{c}
      \sum_{k=1}^l m^k_i = m_i -\ind_{\{1\leq i\leq l\}}, \ \! 1\leq i\leq n \\
      0\leq m^k_i \leq m_{i-1}^k, \ 2\leq i\leq n, \ \! 1\leq k\leq l
  \end{subarray}} \prod_{k=1}^l \frac{a_0( m^k_1,\ldots,m^k_n )}{1 +
     m^k_1+\cdots +m^k_n }
   \\
   & =
     {\P}( N_t = m_1+\cdots +m_n )
     \sum_{l=1}^n \frac{\ind_{\{
      m_l > m_{l+1}
      \}}}{l!} \hskip-0.4cm
   \sum_{\begin{subarray}{c}
      \sum_{k=1}^l m^k_i = m_i -\ind_{\{1\leq i\leq l\}}, \ \! 1\leq i\leq n \\
      0\leq m^k_i \leq m_{i-1}^k, \ 2\leq i\leq n, \ \! 1\leq k\leq l
  \end{subarray}} \prod_{k=1}^l \frac{a_0( m^k_1,\ldots,m^k_n )}{1 +
     m^k_1+\cdots +m^k_n }
\end{align*}
  from
  \eqref{recursion-total-sol}, 
 which yields
 \eqref{recursion-1-coef-general-j} when $j=0$
 and 
 $1\leq m_i \leq m_{i-1}$, $2\leq i\leq n$. 
 
\noindent 
$(ii)$ By iterating \eqref{eqn-4} over
{$n-j-1$} steps {and
then using \eqref{eqn-4-2}}, we obtain
\begin{align*}
  & p_{t,j} ( 
  m_1, \ldots ,{m_j},m_{j+1}+1,m_{j+2},\ldots m_n)
  \\
  & = \ind_{\{m_{j+1} \geq m_{j+2}\}} \lambda e^{-\lambda t} \int_0^t
  p_{s,0}
  ( m_1 , \ldots , m_n) ds
  \\
  &
  \quad
  + \lambda \int_0^t e^{-(t-s)\lambda}
    \\
  &
\quad \sum_{\begin{subarray}{c}
      m'_i + m''_i = m_i - \ind_{\{1\leq i=j+2\}}, \ \! 1\leq i\leq n \\
      0\leq m'_i\leq m'_{i-1}, \ 2\leq i\leq n \\
      0\leq m''_i\leq m''_{i-1}, \ 2\leq i\leq n \atop i\ne j+3 \\
      0\leq m''_{j+3} \leq m''_{j+2}+1
  \end{subarray}} p_{s,0}
  ( m'_1 , \ldots , m'_n )
   {p_{s,j+1}} ( 
   m''_1, \ldots , {m''_{j+1}}, m''_{j+2}+1
   {,} 
   m''_{j+3}, \ldots , m''_n 
   )
 ds \\
 & = \ind_{\{m_{j+1} \geq m_{j+2}\}} \lambda e^{-\lambda t} \int_0^t
 p_{s,0}
  ( m_1, \ldots , m_n ) ds \\
  &\quad + \ind_{\{
    m_{j+2} > m_{j+3}
    \}} \lambda^2 e^{-\lambda t} \int_0^t \int_0^s \sum_{\begin{subarray}{c}
      m^1_i + m^2_i = m_i - \ind_{\{1\leq i= j+2\}}, \ \! 1\leq i\leq n \\
      0\leq m^1_i\leq m^1_{i-1}, \ 2\leq i\leq n \\
      0\leq m^2_i\leq m^2_{i-1}, \ 2\leq i\leq n
 \end{subarray}} p_{s,0}
  ( m^1_1 , \ldots , m^1_n ) 
  p_{r,0} ( 
   m^2_1, \ldots ,m^2_n 
 ) dr ds \\
    &\quad + \lambda^2 \int_0^t \int_0^s e^{(r-t)\lambda} \sum_{\begin{subarray}{c}
      m^1_i + m^2_i + m^3_i = m_i - \ind_{\{j+2\leq i\leq j+3\}}, \ \! 1\leq i\leq n \\
      0\leq m^1_i\leq m^1_{i-1}, \ 2\leq i\leq n \\
      0\leq m^2_i\leq m^2_{i-1}, \ 2\leq i\leq n \\
      0\leq m^3_i\leq m^3_{i-1}, \ 2\leq i\leq n, \ i\ne j+4 \\
      0\leq m^3_{j+4} \leq m^3_{j+3}+1
   \end{subarray}} p_{s,0}
   ( m^1_1 , \ldots , m^1_n ) \\
   &
   \qquad \qquad \qquad \qquad \qquad \qquad 
   p_{r,0}
   (  m^2_1 , \ldots , m^2_n ) {p_{r,j+2}} ( 
   m^3_1,\ldots , {m^3_{j+2}}, m^3_{j+3}+1
   {,} 
   m^3_{j+4},\ldots , m^3_n ) dr ds \\
    & = \cdots \\
  & = e^{-\lambda t}
  \sum_{l=1}^{n-j} \ind_{\{m_{j+l}-\ind_{\{l\ge2\}} \geq m_{j+l+1}\}} \lambda^l  \\
    &\quad \int_{0\leq s_l\leq \cdots \leq s_1\leq t} \sum_{\begin{subarray}{c}
      \sum_{k=1}^l m^k_i = m_i -\ind_{\{j+2\leq i\leq j+l\}}, \ \! 1\leq i\leq n \\
      0\leq m^k_i \leq m_{i-1}^k, \ 2\leq i\leq n, \ \! 1\leq k\leq l
  \end{subarray}} \prod_{k=1}^l p_{s_k,0}
  ( m^k_1 , \ldots , m^k_n ) ds_l \cdots ds_1 \\
  & = e^{-\lambda t}
  (1-e^{-\lambda t})^{1 +
    m_1+\cdots +m_n } \sum_{l=1}^{n-j} \frac{\ind_{\{m_{j+l}-\ind_{\{l\ge2\}} \geq m_{j+l+1}\}}}{l!}
    \hskip-1.2cm
    \sum_{\begin{subarray}{c}
      \sum_{k=1}^l m^k_i = m_i -\ind_{\{j+2\leq i\leq j+l\}}, \ \! 1\leq i\leq n \\
      0\leq m^k_i \leq m_{i-1}^k, \ 2\leq i\leq n, \ \! 1\leq k\leq l
    \end{subarray}} \prod_{k=1}^l \frac{a_0( m^k_1,\ldots,m^k_n )}{1 +
      m^k_1+\cdots +m^k_n }, 
\end{align*}
 from which \eqref{recursion-1-coef-general-j} follows.
\end{Proofy}
\begin{Proofy} {\em of Proposition~\ref{hjkfd21}}. 
 We proceed by induction on $m \geq 0$. 
 We let 
$$
 A_j^\sigma(m)
 := 
 \E_j \left[
            \prod_{k=1}^{{m+j}} 
   \sigma (k)^{X_t^{(k)}}
   \ \! \Big| \ \!
      N_t = m
        \right], \quad j\geq 0, 
 $$
 with ${A}_j^\sigma (0)= 1$. 
 By \eqref{fjklds13}, we have 
$$
 A_j^\sigma(m)
 = 
  \sum_{\begin{subarray}{c}
     m_1+\cdots +m_n = m, \ n\geq 0, \\
      1\leq m_i \leq m_{i-1}, \ 2\leq i\leq n
    \end{subarray}} a_j^\sigma (m_1,\ldots,m_n)
,
$$ 
 where
 $$
 a_j^\sigma (m_1,\ldots,m_n)
 :=
 a_j (m_1,\ldots,m_n)
 \prod_{k=1}^n \sigma(k)^{m_k}, 
$$
 and $\sigma ( k) := \gamma + k -2$, $k\geq 1$. 
 By the induction relation 
 \eqref{recursion-1-coef-general-j},
 similarly to \eqref{fjkld3-0}, we have 
\begin{align}
\nonumber %
   & A^\sigma_j (m+1) = \sum_{\begin{subarray}{c}
     m_1+\cdots +m_n = m+1, \ n\geq 1, \\
      1\leq m_i \leq m_{i-1}, \ 2\leq i\leq n
    \end{subarray}} a^\sigma_j (m_1,\ldots,m_n)
    \\
    \nonumber
  & \quad = \sum_{n=j+1}^{m+j+1}\sum_{\begin{subarray}{c}
    m_1+\cdots +m_n = m+1 \\
    1\leq m_i \leq m_{i-1}, \ 2\leq i\leq n
    \end{subarray}} \sum_{l=1}^{n-j} 
    \frac{1}{l!}
    \ind_{\{
        m_l > m_{l+1}
        \}}
    \sum_{\begin{subarray}{c}
      \sum_{k=1}^l m^k_i = m_i -\ind_{\{1\leq i\leq l\}}, \ \! 1\leq i\leq n \\
      0\leq m^k_i \leq m_{i-1}^k, \ 2\leq i\leq n, \ \! 1\leq k\leq l
    \end{subarray}} \prod_{k=1}^l \frac{a^\sigma_0( m^k_1,\ldots,m^k_n )}{1 +
      m^k_1+\cdots +m^k_n } \\
\nonumber
&\quad = \sum_{l=1}^{m+1}
\frac{1}{l!}
\sum_{n'=1}^{m+1-l} \sum_{\begin{subarray}{c}
      m'_1+\cdots +m'_{n'} = m+1-l \\
      1\leq m'_i \leq m'_{i-1}, \ 2\leq i\leq n'
\end{subarray}}
\sum_{\begin{subarray}{c}
      \sum_{k=1}^l m^k_i = m'_i, \ \! 1\leq i\leq n' \\
      0\leq m^k_i \leq m_{i-1}^k, \ 2\leq i\leq n', \ \! 1\leq k\leq l
\end{subarray}} \prod_{k=1}^l \frac{a^\sigma_0( m^k_1,\ldots,m^k_{n'} )}{1 +
  m^k_1+\cdots +m^k_{n'} }
\\
\nonumber
        &\quad = \sum_{l=1}^{m+1}\frac{1}{l!} \sum_{\begin{subarray}{c}
      \sum_{k=1}^l m_k = m+1-l \\
      m_1,\ldots , m_l \geq 0
    \end{subarray}} \sum_{n'\geq 1}\sum_{\begin{subarray}{c}
      m^k_1+\cdots +m^k_{n'} = m_k, \ \! 1\leq k\leq l \\
      0\leq m^k_i \leq m_{i-1}^k, \ 2\leq i\leq n', \ \! 1\leq k\leq l \\
      \text{at least one of } m^k_{n'}, \ \! 1\leq k\leq l \text{ is nonzero}
\end{subarray}} \prod_{k=1}^l \frac{a^\sigma_0( m^k_1,\ldots,m^k_{n'} )}{m_k+1}
\\
\nonumber
        &\quad = \sum_{l=1}^{m+1} \frac{1}{l!} \sum_{\begin{subarray}{c}
      m_1 + \cdots + m_l = m+1-l \\
      m_1, \ldots , m_l \geq 0
\end{subarray}} \prod_{k=1}^l
\left(
\frac{1}{m_k+1} \sum_{n_k \geq 0} \sum_{\begin{subarray}{c}
      m^k_1+\cdots + m^k_{n_k} = m_k \\
      1\leq m^k_i \leq m_{i-1}^k, \ 2\leq i\leq n_k
    \end{subarray}} a^\sigma_0 ( m^k_1,\ldots,m^k_{n_k} )
\right)
\\
\label{eqn-8}
&\quad = \sum_{l=1}^{m+1}
\frac{1}{l!}
\left( \prod_{k=j+1}^{j+l} \sigma(k) \right)
\sum_{\begin{subarray}{c}
      m_1 + \cdots + m_l = m+1 \\
      m_1\geq 1, \ldots , m_l \geq 1
\end{subarray}} \prod_{k=1}^l \frac{A^\sigma_0 (m_k-1)}{m_k}
    ,
 \qquad
  m\geq 0,
\end{align} 
 where %
 in the third equality we made the change of variables
 $m'_i = m_i - \ind_{\{1\leq i\leq l\}}$. 
 {Plugging} %
 the relation
 $$
 \sigma ( k) = \gamma + k -2, \quad k\geq 1,
 $$ 
 {in \eqref{eqn-8}}, we have 
 $$
  A^\sigma_0 (m+1) 
  = \sum_{l=1}^{m+1}
  \binom{l+\gamma -2}{l}
  \sum_{\begin{subarray}{c}
      m_1 + \cdots + m_l = m+1 \\
      m_1\geq 1, \ldots , m_l \geq 1
\end{subarray}} \prod_{k=1}^l
\left( \frac{
   (-\gamma )^{m_k-1} 
}{m_k}
\binom{
      - 1 
      + 1/{\gamma}
}{m_k-1}
\right)
,
 \quad
  m\geq 0,
$$ 
 and Lemma~\ref{lemma-comb-generalized} then 
 shows that ${A}_0^\gamma (m+1)$
 is the coefficient of $x^{m+1}$ in the series
\begin{align} 
 \nonumber 
  &
 \sum_{l=1}^\infty \binom{l+\gamma -2}{l} \bigg( \sum_{n=1}^\infty 
  \frac{(-\gamma )^{n-1}}{n} \binom{
 -1 + 1 / \gamma  
 }{n-1} x^n \bigg)^l
  = 
  \sum_{l=1}^\infty  (-l)^l \binom{1-\gamma }{l}
  \big(
  1- ( 1- \gamma x )^{1 / \gamma } 
  \big)^l
\\
\nonumber %
  & \quad  = (1 -
(
1- ( 1- \gamma x )^{1 / \gamma } 
) 
)^{1-\gamma } - 1
  \\
\nonumber
& \quad  = (1- \gamma x)^{
  -1 + 1 / \gamma 
} - 1
  \\
\nonumber
& \quad = \sum_{m=1}^\infty
(- \gamma )^m \binom{
    -1 + 1 / \gamma 
    }{m} x^m, 
\end{align}
 which allows us to conclude 
 when $j=0$. 
 When $j\geq 1$,
 we have 
 $$
  A^{{\sigma}}_j (m+1) 
  =
  \sum_{l=1}^{m+1}
  \binom{j+l+\gamma -2}{l}
      \sum_{\begin{subarray}{c}
            m_1 + \cdots + m_l = m+1 \\
      m_1, \ldots ,m_l \geq 1
    \end{subarray}} \prod_{k=1}^l
    \left(
- (- \gamma )^{m_k} \binom{
    1 / \gamma 
    }{m_k}
\right),
 \quad
  m\geq 0,
$$ 
 hence 
 Lemma~\ref{lemma-comb-generalized} shows that, letting
 \begin{equation*}
 Z_\gamma (x): = - \sum_{n=1}^\infty ( - \gamma )^n \binom{ {1} / {\gamma}}{n} x^n = 1- (1- \gamma x)^{{1} / {\gamma}},  
\end{equation*}
 the quantity
 $A_j^{{\sigma}} (m+1)$ is the coefficient of $x^{m+1}$ in the series
\begin{align*}
  \sum_{l=1}^\infty \binom{j+l+\gamma -2}{l}
  (Z_\gamma (x))^l &= \sum_{l=1}^\infty (-l)^l \binom{-(j-1+\gamma)}{l}
  ( Z_\gamma (x) )^l
  \\
  & =
  \frac{1}{(1-Z_\gamma (x))^{j-1+\gamma}} -1 \\
  &= \sum_{m=1}^\infty (- \gamma x )^m \binom{
    -1 - (j-1)/\gamma }{m},
\end{align*}
which yields \eqref{fjkl234}.
\end{Proofy}

\paragraph{Acknowledgement.}
 This research is supported by the National Research Foundation, Singapore. %
 The work of Q. Huang is supported by
 the National Natural Science Foundation of China under Grant No. 12501241,
 the Basic Research Program of Jiangsu under Grant No. BK20251280,
 the Zhishan Young Scholar Program of Southeast University,
 the Start-Up Research Fund of Southeast University under Grant No. RF1028624194, and
 the Jiangsu Provincial Scientific Research Center of Applied Mathematics under Grant No. BK20233002.

\footnotesize

\def\cprime{$'$} \def\polhk#1{\setbox0=\hbox{#1}{\ooalign{\hidewidth
  \lower1.5ex\hbox{`}\hidewidth\crcr\unhbox0}}}
  \def\polhk#1{\setbox0=\hbox{#1}{\ooalign{\hidewidth
  \lower1.5ex\hbox{`}\hidewidth\crcr\unhbox0}}} \def\cprime{$'$}

\end{document}